\documentclass[12pt,onecolumn,journal]{IEEEtran}

\usepackage[T1]{fontenc}
\usepackage{amsmath,amsfonts,amssymb}
\usepackage{algorithm}
\usepackage{algorithmic}
\usepackage{array}
\usepackage{booktabs}
\usepackage[caption=false,font=normalsize,labelfont=rm,textfont=rm]{subfig}
\usepackage{textcomp}
\usepackage{graphicx}
\usepackage{cite}
\usepackage{url}
\usepackage[hidelinks]{hyperref}
\usepackage{orcidlink}
\usepackage{lineno}
\usepackage{setspace}

\makeatletter
\newsavebox{\IEEEsubfigurebox}
\newcommand{\IEEEsubfigurecaption}{}
\newcommand{\IEEEsubfigurelabel}{}
\newenvironment{subfigure}[2][b]{%
  \global\let\IEEEsubfigurecaption\@empty
  \global\let\IEEEsubfigurelabel\@empty
  \begin{lrbox}{\IEEEsubfigurebox}%
    \begin{minipage}[#1]{#2}%
      \renewcommand{\caption}[1]{\gdef\IEEEsubfigurecaption{##1}}%
      \renewcommand{\label}[1]{\gdef\IEEEsubfigurelabel{##1}}%
}{%
    \end{minipage}%
  \end{lrbox}%
  \ifx\IEEEsubfigurecaption\@empty
    \usebox{\IEEEsubfigurebox}%
  \else
    \subfloat[\IEEEsubfigurecaption]{%
      \usebox{\IEEEsubfigurebox}%
      \ifx\IEEEsubfigurelabel\@empty\else
        \label{\IEEEsubfigurelabel}%
      \fi
    }%
  \fi
}
\makeatother

\newcommand{\sep}{\unskip, }
\newenvironment{keywords}{\begin{IEEEkeywords}}{\end{IEEEkeywords}}

\newcommand{\orcidlinkurl}[2]{%
  \begingroup
  \let\savedhref\href
  \renewcommand{\href}[2]{\savedhref{#1}{##2}}%
  \orcidlink{#2}%
  \endgroup
}
\providecommand{\doi}[1]{doi:\ \href{https://doi.org/#1}{\nolinkurl{#1}}}

\providecommand{\balance}{}

\begin{document}
\let\WriteBookmarks\relax

\title{Unsupervised Spatiotemporal Source Separation of Dolphin Whistles in Passive Acoustic Monitoring}

\author{Lixing~Chen\,\orcidlinkurl{https://orcid.org/my-orcid?orcid=0000-0001-5245-8898}{0000-0001-5245-8898}%
  and Roee~Diamant\,\orcidlinkurl{https://orcid.org/0000-0002-2430-7061}{0000-0002-2430-7061},~\IEEEmembership{Senior Member,~IEEE}%
  \thanks{Lixing Chen and Roee Diamant are with the Hatter Department of Marine Technologies, University of Haifa, Haifa, Israel. Roee Diamant is also with the Faculty of Electrical Engineering and Computing, University of Zagreb, Zagreb, Croatia. Corresponding author: Roee Diamant (e-mail: roee.d@univ.haifa.ac.il).}%
  \thanks{This work was sponsored in part by the European Union's Horizon Europe program under the UWIN-LABUST project (project number 101086340)}%
  \thanks{Short running title: Separation of dolphin sources.}%
}

%% JOE review format: 12 pt, single column, double spaced, continuous line
%% numbering, page numbers, and no text in the running header.
\markboth{}{}
\pagestyle{plain}
\doublespacing
\setlength{\emergencystretch}{2em}
\maketitle
\thispagestyle{plain}

%% Manuscript content migrated verbatim from submit_toESWA_anonymous.tex.
\begin{abstract}
Estimating the biomass of dolphins is an essential tool for assessing the health of the marine environment. For long-term analysis, biomass can be evaluated by processing dolphin whistles. However, these whistles often overlap in time, so that a pod of dolphins can generate hundreds of whistles. As a result, it is challenging to estimate the number of dolphins. We present the first unsupervised clustering approach for source separation of dolphin whistles within a small pod of around 10 individuals. Rather than relying on the structure of the signal, our clustering method leverages the spatial separation as reflected in the reception geometry and propagation paths. Specifically, we consider the bearing angle, the time difference of arrival, and the slant delay. The algorithm first performs clustering by fitting a local mixture model to the channel features within short buffers. These buffers are then clustered using the Hungarian assignment framework to maintain consistent source labels over time. Finally, candidate labels for each whistle are fused using a voting system. Our method requires neither a signature-whistle catalog nor species prior and uses only a single user-defined parameter. Our approach scales nearly linearly with the number of whistles, making it suitable for long-term passive acoustic monitoring. Results from the simulations as well as a controlled sea experiment demonstrate a favorable F1-score compared to benchmark clustering schemes, with robustness across motion patterns, measurement noise levels and scalability for the number of sources. Demonstration is also given for real dolphin recordings.
\end{abstract}


\maketitle

\section{Introduction}
\begin{figure}
		\centering
		\includegraphics[width=0.5\columnwidth]{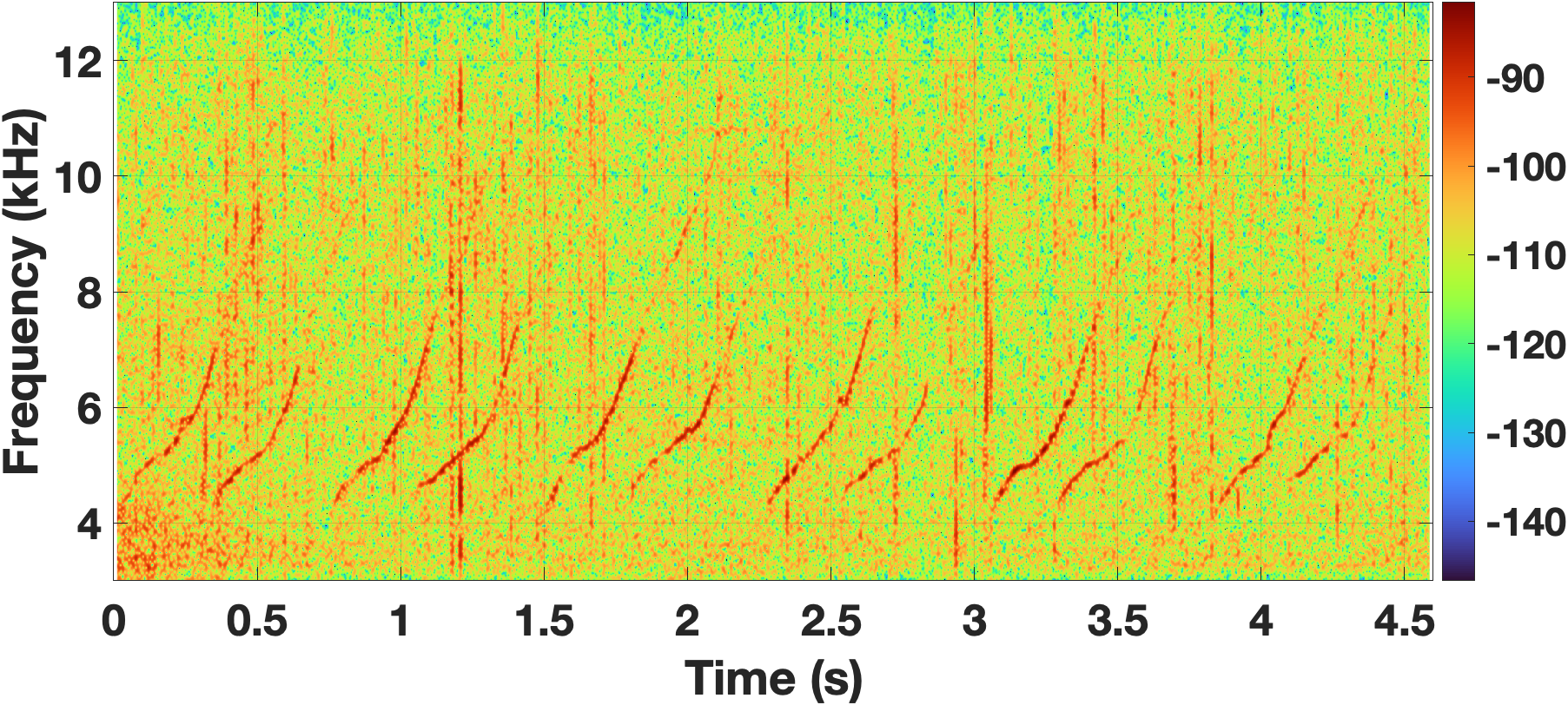}
		\caption{Cocktail-party-like dolphin whistle scene. The spectrogram displays a short $4.6$~s recording segment containing multiple densely interleaved whistles with partially overlapping time--frequency contours. Although several whistle traces are visually detectable, their similar frequency ranges, temporal proximity, and local overlaps make it difficult to determine from the spectrogram alone which whistles were produced by the same dolphin.}
		\label{fig:dolphin_whistleparty}
	\end{figure}

	Presence detection and density distribution of dolphins are important indicators for studying dolphin behavior and their social interactions. As a top predator, the number of dolphins also serves as a measure of the health of the marine environment \cite{mellinger2007overview}. Among the dolphin vocalizations recorded by passive acoustic monitoring (PAM), we focus on the whistle as an important acoustic signal for interactions and thus as a discriminative signal. Studies have shown that dolphin whistles are related to social bonding within a group \cite{papale2020social}, and that the structure of the whistle is closely tied to individual recognition, group association, and behavioral context \cite{janik2013communication,diazlopez2011whistle,akkaya2023initial}. Separating whistles by their source can therefore be a key enabling technique to evaluate the number of dolphins within a pod. Separating dolphin whistles is also important for comparing vocalization activity across individuals, studying group structure and acoustic interaction, and examining the relationship between whistles and behavior. However, the problem is challenging, as dolphins emit frequent whistles in a cocktail party fashion, with hundreds of overlapping signals detected over short time periods, as illustrated in Fig.~\ref{fig:dolphin_whistleparty}. While whistles can be associated with sources by the so-called signature whistles \cite{janik2013identifying,crittenden2025acoustic}, in situ, with no prior information on the individual dolphins or even their species, the time-frequency pattern of the whistle cannot be used as an identification aid.
    
	Separation of dolphin whistles into sources can be considered a clustering problem of detected signals, accounting for simultaneous emissions by multiple dolphins as well as seemingly unorganized patterns in the vocalizations. Unlike, for example, echolocation clicks, which commonly form click trains and whose sources can thus be associated through temporal and rhythmic patterns \cite{lepper2005separation,baggenstoss2011separation,gubnitsky2025source}, dolphin whistles are believed to be independent signals that do not form regular patterns \cite{lopez2020contexts}. Furthermore, the characteristics of dolphin whistles are considered mutual, and no work has yet identified individual-based features within the signal. This is not the case for signature whistles, which do carry information about individual identity and have been widely used for individual recognition and studies of social communication \cite{janik2013communication,janik2013identifying,fearey2019identification}. However, signature whistles are seldom emitted and require prior knowledge of the individuals within the pod. Relying on whistle features alone is arguably insufficient to accomplish unsupervised source separation.
	
	Current approaches to dolphin source separation rely on visual markers, template comparison, and feature analysis. Photo-identification recognizes individuals from notches, scars, pigmentation on the fin, or other visual markings \cite{wursig1977photographic,urian2015recommendations,maglietta2023arianna}. Signature whistle attribution can link a particular whistle type to an already identified individual \cite{janik2013identifying,fearey2019identification,erbe2020matching,crittenden2025acoustic}. Supervised deep learning methods reduce the burden of manual processing and have shown promise in whistle detection, contour extraction and classification of signature whistles \cite{jensen2024automatic,miao2024novel}, and source separation in other species \cite{bermant2021biocppnet}. An interesting source separation technique for the latter is the use of multi-target tracking to associate click sequences by source, but its core statistics assume a repetitive structure of the emitted signals, which does not seem to represent dolphin whistles well \cite{baggenstoss2011separation,gubnitsky2025source}. These approaches face challenges in scenarios where no information is available on dolphin mobility and no prior knowledge is provided on the characteristics of their whistles. In particular, even when a signature whistle of a specific dolphin is identified, it is difficult to assume that such information is available for all dolphins in the pod. 

	Considering these challenges, we propose an unsupervised dolphin whistle source separation framework that, rather than relying on the signals' structure, leverages the spatial separation of dolphins as reflected in the characteristics of the dolphin-to-receiver acoustic channel. Given a set of detected whistle symbols, each whistle is represented by two features: angle, measured by the signal's bearing or time-difference-of-arrival (TDoA), and slant delay, which measures the time delay between the direct path and the first dominant reflection path. To address feature drift caused by dolphin movement and changes in the channel impulse response, we use a two-step approach: clustering is first performed on short time buffers, which are then combined for full association within the observation window. Within each buffer, a local mixture model forms short-term source hypotheses. The local clusters are then aligned between adjacent buffers using the Hungarian assignment framework to fuse source associations in the overlapping regions of the buffers.
	Our contribution is threefold:
	\begin{enumerate}
		\item The first blind source separation of dolphin whistles by clustering channel features.
		\item A computationally efficient clustering solution that manages temporal changes in the channel through buffer-to-buffer source association.
		\item A cross-buffer tracking solution that exploits the temporal continuity of features to link local clusters and fuse overlapping buffers into a source labeling consistent over time.
	\end{enumerate}
	We evaluate the performance of our clustering approach through numerical simulations and controlled sea experiments. Compared to baseline clustering methods, our results show a favorable precision-recall tradeoff under challenging measurement noise conditions. Scalability is demonstrated with respect to the number of sources, and robustness is shown across various dynamic patterns, different sea environments, and the choice of the single user-defined parameter we use. Clustering results from real recordings of dolphin whistles demonstrate the applicability of our approach to realistic scenarios. We share our code for reproducibility.
	
	The remainder of this paper is organized as follows. Section~\ref{sec:related_work} surveys related work on the challenge of dolphin whistle clustering. Section~\ref{sec:system_model} introduces our system model, main assumptions, and preliminaries regarding dolphin whistles. Section~\ref{sec:method} details our proposed clustering method and provides a complexity analysis. Section~\ref{sec:results} presents simulation results, findings from a controlled experiment, and analysis of real recordings of dolphin whistles. Section~\ref{sec:discussion} discusses the advantages and limitations of our work, and Section~\ref{sec:conclusions} presents our conclusions.

	\section{Related Work}
	\label{sec:related_work}
    The common practice for dolphin source separation relies on visual photo-identification, determining pod size and individual identity from unique fin notches and marks \cite{wursig1977photographic}. Because this requires marked individuals and annotated databases, it is commonly supplemented by acoustic source identification using signature whistles.
	Dolphins' signature whistles are believed to serve as unique identifiers for each dolphin. As such, they are used to estimate the number of dolphins in a pod. Identification of dolphins' signature whistles is proposed in \cite{janik2013identifying} by matching the whistle's contour through visual categorization of repeated bouts of the same contour type within signature inter-whistle intervals. 
    To increase confidence in whistle signature identification, the methods in \cite{erbe2020matching,crittenden2025acoustic} combine acoustic and visual information. This is done by associating repeated signature-whistle contours with photo-identified individuals from the same encounters. However, this solution is relevant only to a very limited number of recognized individuals. When a local database of signature whistles is not available, machine learning classification approaches have been proposed.
	
	Existing supervised machine learning methods for whistle classification rely on extracting acoustic features from whistles. A common approach is to extract the time-frequency contour of each detected whistle and then cluster or classify it. 
    The identified contour serves as the basis for clustering. To address the challenges of multipath and low SNR, the method in \cite{kipnis2021graph} clusters the dominant whistle contour, whistle fragments, their harmonics, and their multipath. This method applies weighting based on the structural match of the whistle within a single graph Laplacian utility function. However, it does not provide separation based on the whistle's source. A convolutional deep neural network is trained to learn contour patterns from the large Sarasota Dolphin whistle database for segmenting signature whistles of bottlenose dolphins \cite{jensen2024automatic}. Unique separation of whistles is assumed. For the common cocktail-party problem, where multiple whistles overlap, the training in \cite{bermant2021biocppnet} mixes and shifts labeled individual signals to train a lightweight U-Net for source separation. Although this method has shown potential for biological acoustic source separation, the need for labeled data prevents its application in in-situ data analysis for unknown pods of dolphins. For these cases, solutions based on sequence tracking are proposed.

	Multi-target tracking (MTT) methods exploit the temporal structure of click sequences for source association. In \cite{lepper2005separation} and \cite{starkhammar2011separating}, changes in amplitude, delays between symbols, and spectral correlation are used to separate overlapping sequences of dolphin clicks from multiple individuals. A linear separation by threshold is considered, which may lead to classification errors in cases of overlapping signals or when estimation errors occur due to noise. 
    Addressing the challenge of an unknown number of sources, \cite{gerard2008feature} employed a multi-hypothesis tracking algorithm to maintain multiple possible associations, thereby achieving higher separation accuracy. This, however, comes at the cost of increased computational complexity and the need for hypothesis pruning or simplification. For long in-situ recordings, \cite{gubnitsky2025source} utilizes short-term spectral and propagation-related information for clustering, then uses sequential structure and source consistency over time to achieve real-time separation of click sounds. However, the application to dolphin whistle separation is not direct, since the acoustic symbol is significantly different, as are the source-assignment statistics, mostly because the whistle symbol does not provide a train-level distribution as a click train does.
	
	While the above approaches have been successfully used to separate dolphin clicks and the vocalizations of other marine cetaceans, these methods assume that the clustered signals have distinct characteristics. This may not be the case for dolphin whistles, which, apart from signature whistles, do not contain distinct source characteristics. We argue that this is why, to the best of our knowledge, no solution has yet been offered for the unsupervised clustering of dolphin whistles by their source. The challenge is to determine the source of each signal in a sequence of simultaneously emitted whistles by several dolphins, with no prior knowledge of their signature whistles, specific features, or even the dolphin species.
	
	\section{System Model}
	\label{sec:system_model}
	\subsection{Assumptions and Setup}
	
	Our setup includes a directional recorder, either a vector hydrophone or an array of at least two hydrophones. The vector hydrophone determines direction by simultaneously capturing acoustic pressure and particle velocity signals at the same spatial point in orthogonal directions. The hydrophone array provides geometric constraints through the time difference of arrival between different receiving channels. The system can provide directional or spatial information during monitoring for feature extraction. The sensors are assumed to be deployed at a depth that allows separation of the direct and primary multipath for slant delay measurements. The received whistles are assumed to propagate with constant sound speed. We claim that this is an acceptable assumption because dolphins tend to interact in the upper layer of the water column \cite{klatsky2007offshore}. 
	
\begin{figure*}
		\centering
		\begin{subfigure}[b]{0.32\textwidth}
			\centering
			\includegraphics[width=\textwidth]{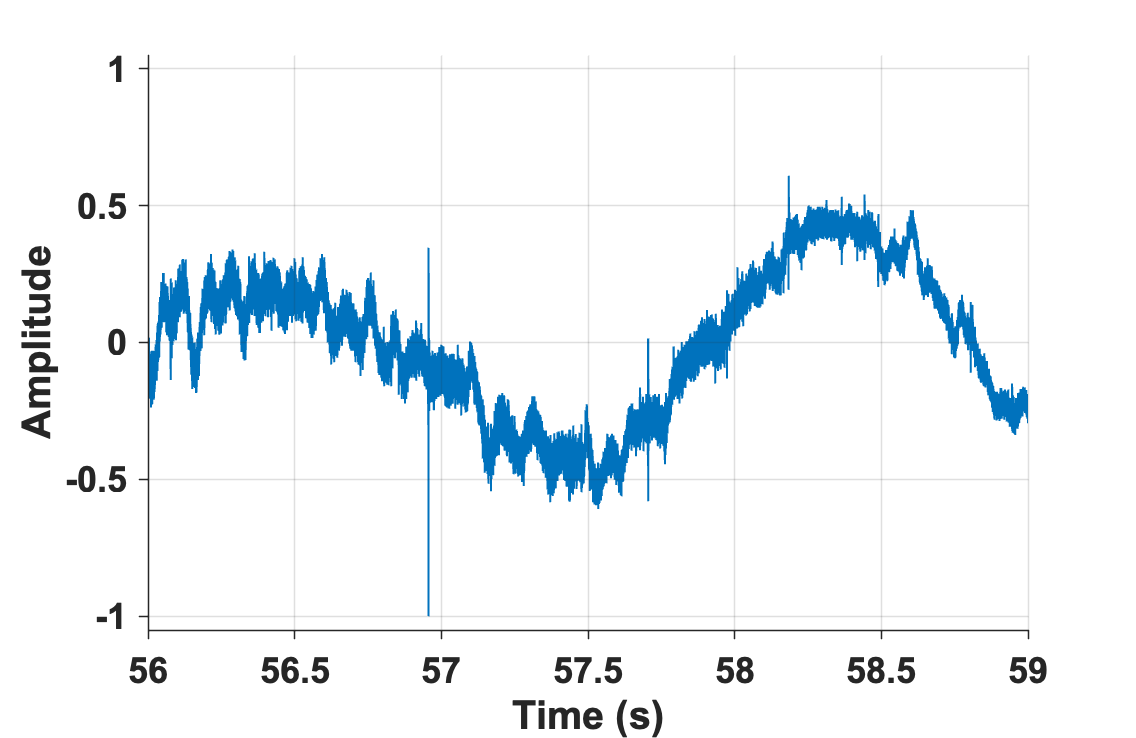}
			\caption{ }
			\label{fig:whistle_waveform}
		\end{subfigure}
		\begin{subfigure}[b]{0.32\textwidth}
			\centering
			\includegraphics[width=\textwidth]{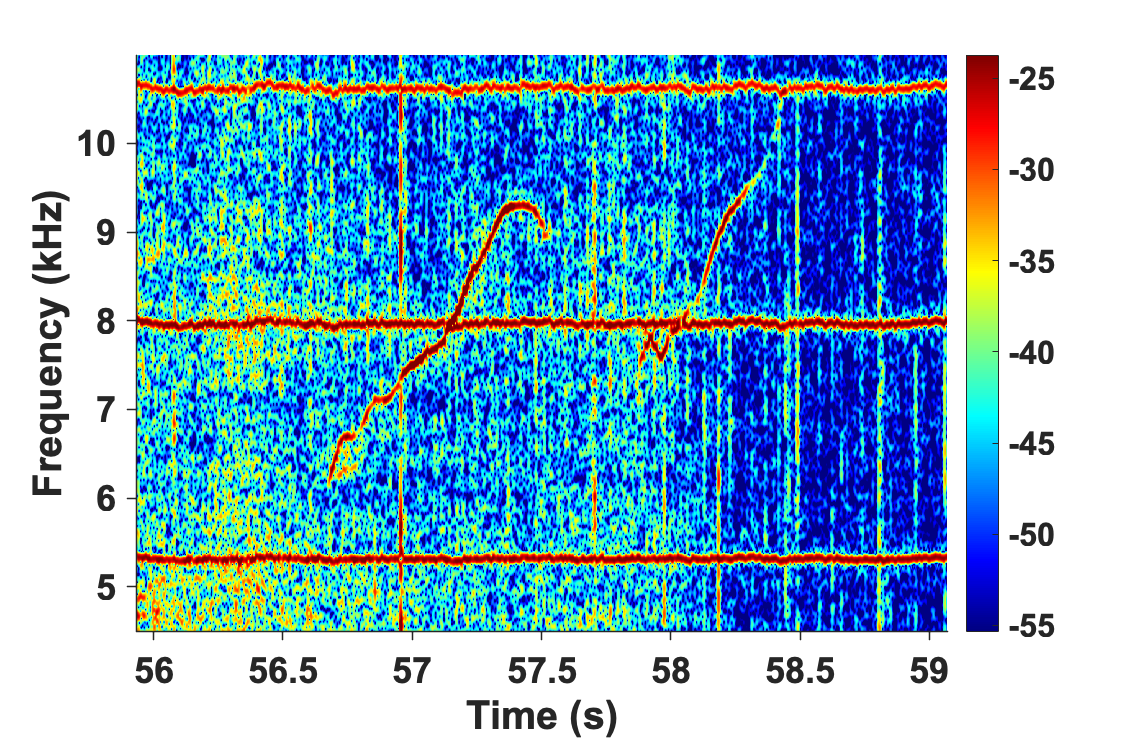}
			\caption{ }
			\label{fig:whistle_spectrogram}
		\end{subfigure}
		\begin{subfigure}[b]{0.32\textwidth}
			\centering
			\includegraphics[width=\textwidth]{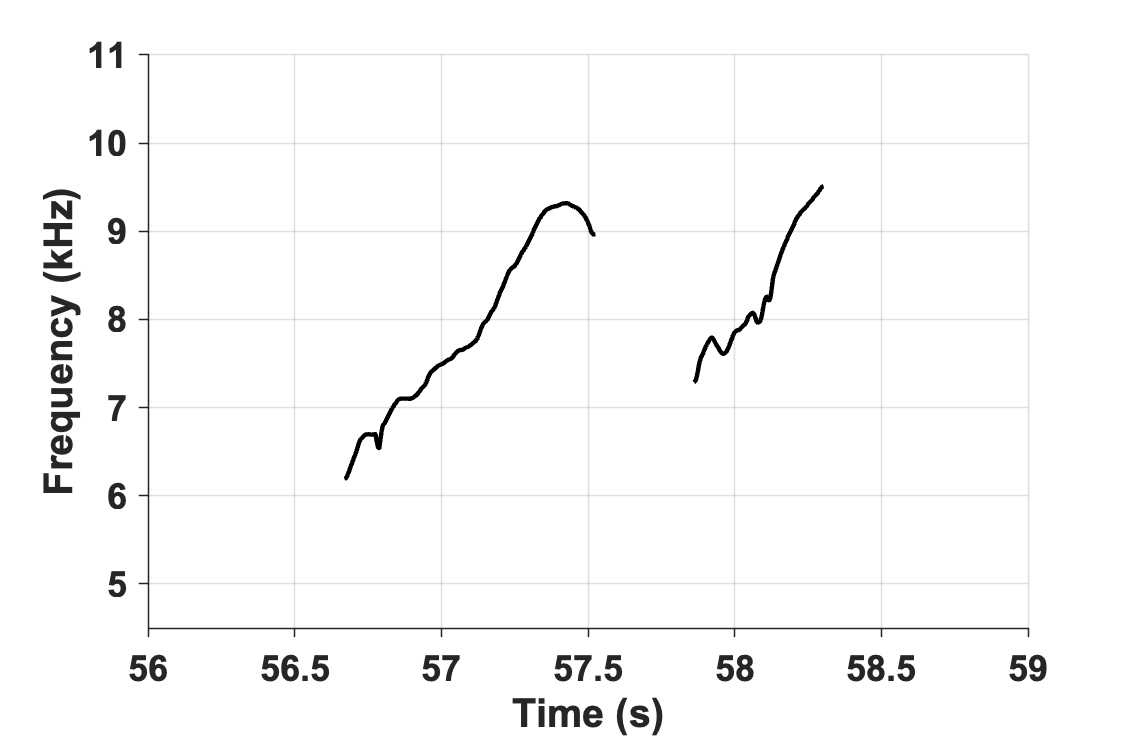}
			\caption{ }
			\label{fig:extracted_whistle_spectrogram}
		\end{subfigure}
		\caption{Example pre-processing of real dolphin whistles recorded by a vector sensor at the Eilat Dolphin Reef, Red Sea. Panel (a) is the time-domain pressure measurement, (b) the corresponding spectrogram, and (c) the whistle contours extracted from the spectrogram. The extracted contours in (c) constitute the detected whistle symbols.}
		\label{fig:real_dolphin_whistle}
	\end{figure*}

	In this paper, whistle source separation does not begin by processing a continuous raw recording stream frame by frame, but rather with the detected whistle sounds. That is, we assume the availability of candidate whistle sound onsets provided by a preliminary detector and annotator, such as in \cite{korkmaz2023automated,roch2011automated}. Each detection is assumed to provide an accurate symbol timestamp for each whistle. We also assume preliminary recognition of the whistle contour, for example by the process in \cite{gruden2020automated,li2023using}, as well as clustering of the whistle contour into harmonics and multipath, for example by the process in \cite{kipnis2021graph}. An example of such a pre-processed procedure over a frame including multiple dolphin whistles is provided in Fig.~\ref{fig:real_dolphin_whistle}. The algorithm provided will show sensitivity to outliers due to false detection of whistles. That said, while these procedures are essential for our source separation algorithm, we do not limit the process to specific detectors or whistle identifiers. 
	
	We introduce a local stability assumption regarding the propagation characteristics of dolphin signals. The features of whistles emitted by the same dolphin within a short time buffer should display local consistency in their propagation and reception geometry. This assumption is based on two arguments. First, dolphin movement is continuous, and its geometry relative to the receiver does not change arbitrarily within a short time period. Second, the direction and propagation-related observations of the same source are not expected to change significantly within a short time window. Although the distribution of features may drift over time due to changes in dolphin movement, depth, and channel conditions, we assume that these features change slowly within an observation window of a few minutes. We also assume that the number of sources within each time buffer can be upper bounded by an unknown value $K$. However, this assumption can be relaxed by comparing source separation iterations for different values of $K$.
	
	We refer to dolphin whistle separation as a clustering problem, where each detected whistle is associated with a single source cluster. 
	Let $\mathcal{X}=\{x_i\}_{i=1}^{N}$ denote a set of $N$ detected whistle symbols in a record, where $x_i$ represents the $i$-th whistle symbol. Each symbol is sorted by its start time or center time and can be represented as $x_i = (t_i,\mathbf{y}_i)$, where $t_i$ is the symbol timestamp and $\mathbf{y}_i$ is the whistle waveform. Each detected whistle symbol is assumed to be emitted by a single dolphin, and consecutive whistles from the same dolphin do not overlap. Let $s_i \in \{1,\ldots, K\}$ denote the unknown source label of symbol $x_i$. The result of the source separation is the label set $\mathbf{S}_{1:N}=\{s_i\}_{i=1}^{N}$.

	\subsection{Preliminaries}
	
	Dolphin whistles are wideband acoustic signals with a chirplet structure, referred to here as \textit{symbols}. Fig.~\ref{fig:whistle_waveform} shows an example of a dolphin whistle time-frequency spectrogram recorded by a vector sensor at the Eilat Dolphin Reef in the Red Sea in January 2026. The figure displays multiple consecutive tonal whistles within approximately two seconds. Each whistle exhibits a harmonic structure with multipath and a continuous contour with a smooth rise in frequency. Weak background noise and dolphin clicks are also present. The identification of whistle contours, as shown in Fig.~\ref{fig:extracted_whistle_spectrogram}, represents the detected whistles.

	The frequency band and duration of dolphin whistles show considerable variation and are influenced by species, population, geographic region, environmental noise, group composition, and behavioral context. Janik and Sayigh's review \cite{janik2013communication} indicates that bottlenose dolphin signature whistles typically span approximately $1$--$30$~kHz, with durations ranging from $0.1$ to $4$~s. Akkaya et al. \cite{akkaya2023initial} also reported bottlenose dolphin whistles with frequencies covering $659$~Hz to $26.122$~kHz and durations of approximately $500$–$570$~ms. Due to these variations in the literature, we avoid relying on these boundaries for separation beyond the basic bandpass filter at the initial processing step. The dolphin whistle is represented as a time-frequency varying symbol with a main carrier frequency \cite{gruden2016automated}.

	A dolphin whistle is characterized by its duration, minimum frequency, maximum frequency, number of maxima, number of saddle points, start frequency, end frequency, peak frequency, central frequency, frequency range, mean or median frequency, contour slope, number of inflection points, number of steps, number of loops, and the presence or absence of harmonics \cite{akkaya2023initial,oswald2007tool,kaplan2017whistle}. These parameters represent the time-frequency function of the whistle and are therefore widely used for species comparison, signature whistle identification, population comparison, and behavioral context analysis \cite{diazlopez2011whistle,akkaya2023initial,kaplan2017whistle}. However, the large variation of these features in the whistles of a single dolphin does not permit their use for source separation. For these reasons, in this work we do not rely on the symbol's features to determine the source dolphin and instead consider the channel's characteristics as a discriminative metric.

	\section{The Dolphin Whistle Clustering Method}
	\label{sec:method}
	\subsection{Key Idea}
\begin{figure*}
		\centering
		\begin{subfigure}[b]{1\textwidth}
			\centering
			\includegraphics[width=\textwidth]{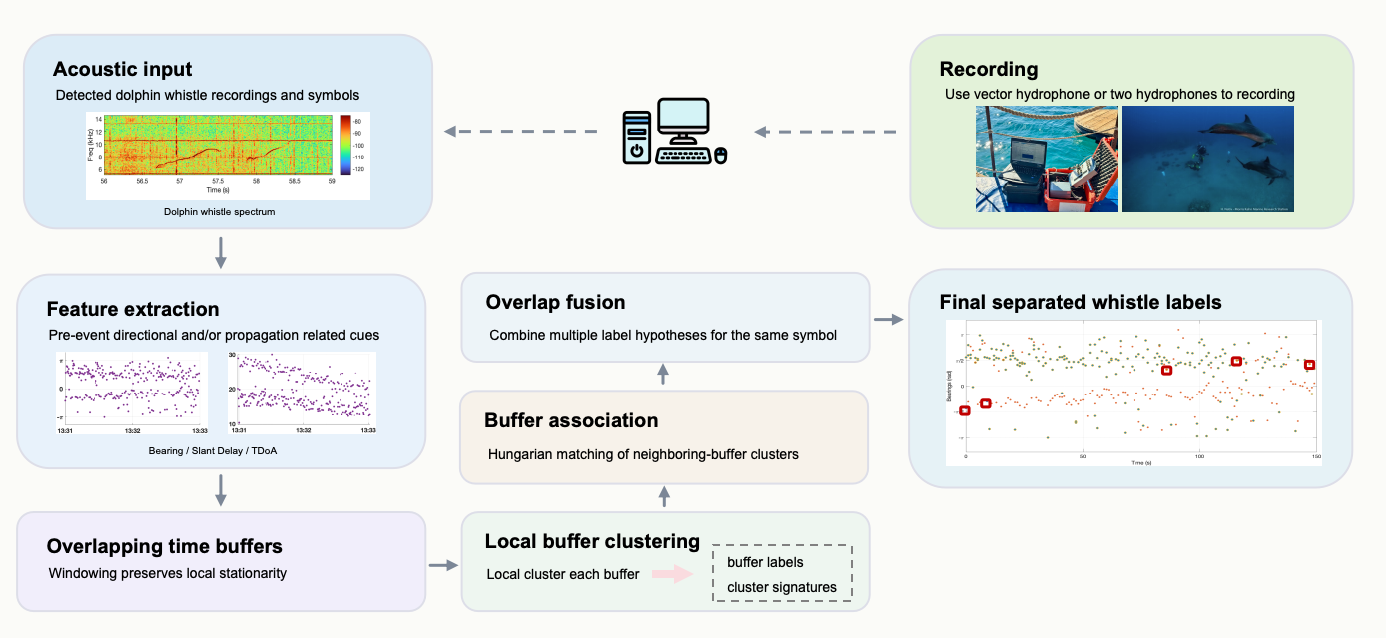}
		\end{subfigure}
    		\caption{The diagram of the proposed source-separation framework. Detected whistle symbols and their recordings, acquired by a vector hydrophone or a two-element array, are first converted into cues, i.e., bearing or TDoA and slant delay. The symbols are then split into overlapping time buffers that preserve local stationarity, and each buffer is clustered independently into short-term source hypotheses. Neighboring-buffer clusters are aligned to build temporally consistent global labels, and the multiple label hypotheses covering each symbol are combined through overlap fusion to produce the final per-whistle source labels.}
		\label{fig:proposed_diagram}
	\end{figure*}

	The key idea of our proposed method is the combination of clustering and tracking in the whistle association. The diagram of our proposed source separation method is shown in Fig.~\ref{fig:proposed_diagram}.
	Buffer-wise local clustering, cross-buffer assignment, and overlap label fusion are included in the proposed method. First, for each detected whistle symbol, the algorithm extracts features. The whistle symbols, arranged in time order, are then divided into a series of overlapping buffers. The algorithm performs local separation only within this short time window, obtaining local labels. The output at this stage is the local partition of whistles within each buffer.
	
	The second step is cross-buffer assignment. Since each buffer forms local clusters independently, the same local numerical label in two neighboring buffers may not correspond to the same source. The cluster association between adjacent valid buffers is formulated as an assignment problem. For a previously aligned source hypothesis in the previous buffer and a local cluster in the current buffer, the algorithm constructs the association cost based on feature differences, such as bearing differences. Hungarian alignment is used to solve the minimum total cost matching, thereby obtaining the local-to-global label mapping for the current buffer. The purpose of this step is to connect the independent local partitions into a temporally continuous source-label structure. The third step is overlap fusion. For each symbol covered by multiple buffers, each symbol receives a set of candidate global labels, one from each covering buffer. These candidates are then fused by reliability-weighted voting, and the label with the largest accumulated support is selected as the final source label. In the following, we describe in detail the different parts of the proposed algorithm. We openly share the processing code in \url{https://anonymous.4open.science/r/dolphin_separation_by_source-9D32/}.

	\subsection{Buffer-wise Local Clustering}
	Accounting for the assumed slow changes in the slant delay and bearing direction of each source dolphin, we avoid batch segmentation and instead cluster whistles within buffers. Recall that the $N$ whistle symbols, sorted in temporal order, are denoted by $\mathcal{X}=\{x_i\}_{i=1}^N$ and the feature vector of each symbol is denoted by $\mathbf{z}_i$. When the system supports bearing estimation, for example, through a vector sensor, we consider $\mathbf{z}_i = (\theta_i,\delta_i)$, where $\theta_i \in [-\pi,\pi)$ is the bearing and $\delta_i \in \mathbb{R}_+$ is the slant delay. When only a two-element small array is available, we take $\mathbf{z}_i = (\tau_i,\delta_i)$, where $\tau_i \in \mathbb{R}$ is the time difference of arrival (TDoA). The detailed estimation of these three features is given in Section~\ref{sec:feature_extraction}. Given the window length $T_\mathrm{buf}$ and the overlap $T_\mathrm{ov} \in [0, T_\mathrm{buf})$, the hop size is $T_\mathrm{hop} = T_\mathrm{buf} - T_\mathrm{ov}$. The $b$-th buffer $\mathcal{B}_b$, the index set $\mathcal{I}_b$ of whistles falling inside it, and its cardinality $N_b$ are
	\begin{equation}
		\begin{aligned}
			&\mathcal{B}_b=[\,t_1+(b{-}1)T_\mathrm{hop},\;t_1+(b{-}1)T_\mathrm{hop}+T_\mathrm{buf}\,], \\
			&\mathcal{I}_b=\{i:t_i\in\mathcal{B}_b\},\quad N_b=|\mathcal{I}_b|.
		\end{aligned}
		\label{eq:buffer_definition}
	\end{equation}
	
	Since $\theta_i$ is a circular variable, we use the von Mises distribution for the bearing and a Gaussian distribution for the slant delay, which are assumed to be conditionally independent given the mixture component. With $J$ candidate components in buffer $b$, the per-component distribution parameters for component $j$ are
	\begin{equation}
		\boldsymbol{\eta}_{b,j}=\bigl(\mu^{\theta}_{b,j},\,\kappa_{b,j},\,\mu^{\delta}_{b,j},\,\sigma^{\delta 2}_{b,j}\bigr),
		\label{eq:local_component_parameters}
	\end{equation}
	where $\mu^{\theta}_{b,j}\in[-\pi,\pi)$ and $\kappa_{b,j}>0$ are the mean direction and concentration of the bearing, and $\mu^{\delta}_{b,j}\in\mathbb{R}$ and $\sigma^{\delta 2}_{b,j}>0$ are the mean and variance of the slant delay. Together with the mixing weights $\pi_{b,j}\ge 0$ satisfying $\sum_{j=1}^{J}\pi_{b,j}=1$, the full mixture parameter set is $\boldsymbol{\Theta}_{b,J}=\{(\pi_{b,j},\boldsymbol{\eta}_{b,j})\}_{j=1}^{J}$, and the local log-likelihood is
	\begin{equation}
		\begin{aligned}
			&\mathcal{L}_b(\boldsymbol{\Theta}_{b,J}) \\
			&=\sum_{i\in\mathcal{I}_b}\log\sum_{j=1}^{J}\pi_{b,j}\,f_\mathrm{vM}(\theta_i;\mu^{\theta}_{b,j},\kappa_{b,j})\,\mathcal{N}(\delta_i;\mu^{\delta}_{b,j},\sigma^{\delta 2}_{b,j}),\\
			&f_\mathrm{vM}(\theta;\mu,\kappa)=\frac{e^{\kappa\cos(\theta-\mu)}}{2\pi I_0(\kappa)},
		\end{aligned}
		\label{eq:local_mixture_density}
	\end{equation}
	where $I_0(\cdot)$ is the modified Bessel function of the first kind of order zero. For a given candidate order $J$, the local fitting problem is the maximum-likelihood estimate
	\begin{equation}
		\hat{\boldsymbol{\Theta}}_{b,J}=\arg\max_{\boldsymbol{\Theta}_{b,J}}\,\mathcal{L}_b(\boldsymbol{\Theta}_{b,J}),
		\label{eq:local_mle}
	\end{equation}
	followed by an order-selection step that returns the local model order $K_b$ together with hard labels derived from the posterior responsibilities.
	
	For each candidate $J$, we solve \eqref{eq:local_mle} using the standard Expectation--Maximization (EM) algorithm \cite{moon1996expectation}. Closed-form updates are available for the mixing weights and slant delay parameters; the bearing mean is obtained from responsibility-weighted circular sums, and the concentration $\kappa_{b,j}$ is recovered from the von Mises moment equation $A(\kappa)=\bar{R}$, where $A(\kappa) = I_1(\kappa)/I_0(\kappa)$ and $\bar{R}$ is the weighted resultant length, solved numerically \cite{mardia2000directional,berens2009circstat}. The variance lower bound $\sigma^{\delta 2}_{b,j} \ge 10^{-4} s^2_{\delta,b}$ prevents degeneracy, and multiple random initializations are used to select the highest-likelihood solution.
	
	Each component has four distribution parameters in $\boldsymbol{\eta}_{b,j}$ and $J{-}1$ independent mixing weights, so the total number of free parameters is $p(J) = 5J - 1$. Requiring at least five symbols per component yields the data-adaptive order bound
	\begin{equation}
		K_{\max,b}=\min\bigl(K_\mathrm{max}^{(g)},\,\lfloor N_b/5\rfloor\bigr),
		\label{eq:adaptive_kmax}
	\end{equation}
	where $K_\mathrm{max}^{(g)}$ is a conservative global cap, and we use $K_\mathrm{max}^{(g)}=15$. The local order is then selected by the Bayesian information criterion (BIC) \cite{schwarz1978estimating},
	\begin{equation}
		K_b=\arg\min_{J\in\{1,\dots,K_{\max,b}\}}\!\Bigl[-2\,\mathcal{L}_b(\hat{\boldsymbol{\Theta}}_{b,J})+(5J-1)\log N_b\Bigr].
		\label{eq:bic_order_selection}
	\end{equation}
	
	Denote by $\beta_{i,j}^{(b)}$ the converged responsibilities under the selected order $K_b$, where the component index now runs over $j=1,\dots,K_b$. The local hard label is the maximum-a-posteriori assignment
	\begin{equation}
		\centering
		\ell_i^{(b)}=\arg\max_{j}\beta_{i,j}^{(b)}, 
	\end{equation}
	forming the local partition $\mathcal{P}_b=\{\mathcal{C}_{b,k}\}_{k=1}^{K_b}$ with $\mathcal{C}_{b,k}=\{i\in\mathcal{I}_b:\ell_i^{(b)}=k\}$. Each local cluster is summarized by a cluster signature
	\begin{equation}
		\mathbf{h}_{b,k}=\bigl(n_{b,k},\,\bar{\theta}_{b,k},\,\hat{\kappa}_{b,k},\,\bar{\delta}_{b,k},\,s^{2}_{\delta,b,k}\bigr),
		\label{eq:cluster_signature}
	\end{equation}
	whose entries are the cluster size $n_{b,k}=|\mathcal{C}_{b,k}|$, the empirical bearing direction and concentration $(\bar{\theta}_{b,k},\hat{\kappa}_{b,k})$, and the empirical slant delay mean and variance $(\bar{\delta}_{b,k},s^{2}_{\delta,b,k})$. They are the sample equivalents of the component parameters and computed directly from the cluster members:
	\begin{equation}
		\label{eq:cluster_signature_stats}
		\begin{aligned}
			&\bar{\theta}_{b,k} = \mathrm{atan2}\!\Bigl(\textstyle\sum_{i\in\mathcal{C}_{b,k}}\sin\theta_i,\;\sum_{i\in\mathcal{C}_{b,k}}\cos\theta_i\Bigr),  \\
			&A(\hat{\kappa}_{b,k}) = \bar{R}_{b,k}, \quad \bar{\delta}_{b,k} = \frac{1}{n_{b,k}}\!\sum_{i\in\mathcal{C}_{b,k}}\!\delta_i,  \\
			&s^{2}_{\delta,b,k} = \frac{1}{n_{b,k}}\!\sum_{i\in\mathcal{C}_{b,k}}\!(\delta_i-\bar{\delta}_{b,k})^{2} 
		\end{aligned}
	\end{equation}
	where $\mathrm{atan2}(y,x)$ is the four-quadrant inverse tangent that maps a Cartesian pair $(x,y)$ to the angle of its vector representation in $[-\pi,\pi)$ while preserving quadrant information, 
	\begin{equation}
		\centering 
        \bar{R}_{b,k}=n_{b,k}^{-1}\sum_{i\in\mathcal{C}_{b,k}}\!e^{\theta_i}
	\end{equation}
	is the cluster-level resultant length and $\hat{\kappa}_{b,k}$
	is recovered from $A(\hat{\kappa}_{b,k})=\bar{R}_{b,k}$ using the same numerical inversion as in EM. The signature $\mathbf{h}_{b,k}$ in \eqref{eq:cluster_signature} is the quantity used in the next stage. Once whistles are clustered within buffers, we proceed with tracking by cross-buffer segmentation.

	\subsection{Temporal Tracking}
	
	\subsubsection{Cross-Buffer Assignment}
	
	A buffer is called \emph{valid} if it contains at least one component, i.e., $K_{\max,b} \ge 1$ (equivalently, $N_b \ge 5$). Buffers with $N_b < 5$ are invalid and skipped, so tracking is performed only over the ordered set of valid buffers. Let $b^{-}$ denote the nearest valid predecessor of buffer $b$, and let $\mathbf{h}_{b^{-}\!,p}$ and $\mathbf{h}_{b,q}$ be two cluster signatures from the two buffers. We introduce the binary decision variable $a_{p,q}^{(b)} \in \{0,1\}$, where $a_{p,q}^{(b)} = 1$ if and only if $\mathcal{C}_{b^{-}\!,p}$ and $\mathcal{C}_{b,q}$ are declared to belong to the same physical source. Because each cluster can be matched to at most one cluster on the other side, the admissible decisions are
	{\small
		\begin{equation}
			\mathcal{A}_b=\Bigl\{\mathbf{A}^{(b)}:\textstyle\sum_{q}a_{p,q}^{(b)}\le 1\;\forall p,\;\sum_{p}a_{p,q}^{(b)}\le 1\;\forall q,\;a_{p,q}^{(b)}\in\{0,1\}\Bigr\}.
			\label{eq:assignment_constraint_set}
		\end{equation}
	}
	
	The empirical centers $(\bar{\theta}_{b,k},\bar{\delta}_{b,k})$ in $\mathbf{h}_{b,k}$ are large-sample estimates with variances
	\begin{equation}
		\begin{aligned}
			&v_{\theta,b,k}=\frac{1}{n_{b,k}\,\hat{\kappa}_{b,k}\,A(\hat{\kappa}_{b,k})}, \\
			&v_{\delta,b,k}=\frac{s^{2}_{\delta,b,k}}{n_{b,k}},
		\end{aligned}
		\label{eq:cluster_center_variance}
	\end{equation}
	where the bearing mean expression follows from the von Mises Fisher information $n\kappa A(\kappa)$ \cite{mardia2000directional}. If the two clusters are independent estimates of the same source center, the residuals
	\begin{equation}
		\begin{aligned}
			&\Delta\theta_{p,q}^{(b)}=\mathrm{wrap}_{[-\pi,\pi)}\!\bigl(\bar{\theta}_{b,q}-\bar{\theta}_{b^{-}\!,p}\bigr),\\
			&\Delta\delta_{p,q}^{(b)}=\bar{\delta}_{b,q}-\bar{\delta}_{b^{-}\!,p},
		\end{aligned}
		\label{eq:transition_residuals}
	\end{equation}
	have zero mean and variances $v_{\theta,p,q}^{(b)}=v_{\theta,b^{-}\!,p}+v_{\theta,b,q}$ and $v_{\delta,p,q}^{(b)}=v_{\delta,b^{-}\!,p}+v_{\delta,b,q}$. The bearing residual is modeled as a zero-mean von Mises with matched concentration $\kappa_{p,q}^{(b)}\triangleq 1/v_{\theta,p,q}^{(b)}$, giving the same source residual likelihood
	\begin{equation}
		\begin{aligned}
			P&\bigl(\Delta\theta_{p,q}^{(b)},\Delta\delta_{p,q}^{(b)}\mid a_{p,q}^{(b)}{=}1\bigr)\\
			&=f_\mathrm{vM}\bigl(\Delta\theta_{p,q}^{(b)};\,0,\,\kappa_{p,q}^{(b)}\bigr)\,\mathcal{N}\bigl(\Delta\delta_{p,q}^{(b)};\,0,\,v_{\delta,p,q}^{(b)}\bigr).
		\end{aligned}
		\label{eq:transition_likelihood}
	\end{equation}
	Taking the negative log-likelihood as a pairwise cost, the cross-buffer assignment is the constrained maximum-likelihood problem
	\begin{equation}
		\begin{aligned}
			&\hat{\mathbf{A}}^{(b)}=\arg\min_{\mathbf{A}^{(b)}\in\mathcal{A}_b}\sum_{p,q}c_{p,q}^{(b)}\,a_{p,q}^{(b)}, \\
			&c_{p,q}^{(b)}=-\log P\bigl(\Delta\theta_{p,q}^{(b)},\Delta\delta_{p,q}^{(b)}\mid a_{p,q}^{(b)}{=}1\bigr).
		\end{aligned}
		\label{eq:cross_buffer_map}
	\end{equation}
	
	To solve ~\eqref{eq:cross_buffer_map}, two issues remain: implausible pairs must be excluded, and the trivial empty solution $\mathbf{A}^{(b)}=\mathbf{0}$ must be ruled out. We address both with a $\chi^2$ gate followed by a maximum-cardinality, minimum-cost matching. Under $a_{p,q}^{(b)}=1$, the standardized squared distance
	\begin{equation}
		\begin{aligned}
			Q_{p,q}^{(b)}&=2\kappa_{p,q}^{(b)}\bigl[1-\cos\Delta\theta_{p,q}^{(b)}\bigr]+\frac{(\Delta\delta_{p,q}^{(b)})^{2}}{v_{\delta,p,q}^{(b)}} \\
			&\;\overset{\Delta\theta\to 0}{\approx}\;\frac{(\Delta\theta_{p,q}^{(b)})^{2}}{v_{\theta,p,q}^{(b)}}+\frac{(\Delta\delta_{p,q}^{(b)})^{2}}{v_{\delta,p,q}^{(b)}}.
		\end{aligned}
		\label{eq:normalized_transition_distance}
	\end{equation}
	is asymptotically $\chi^{2}_{2}$ \cite{mardia2000directional}. Pairs failing the test at level $\alpha=0.05$ are excluded, defining the feasible set
	\begin{equation}
		\mathcal{E}_b=\bigl\{(p,q):\,Q_{p,q}^{(b)}\le\chi^{2}_{2}(0.95)=5.991\bigr\}.
		\label{eq:feasible_transition_set}
	\end{equation}
	Since each cluster contains as few as five samples, the Fisher information variances in \eqref{eq:cluster_center_variance} and the $\chi^{2}_{2}$ distribution in \eqref{eq:normalized_transition_distance} are asymptotic approximations. The matching is then required to use as many feasible pairs as possible before minimizing the total cost:
	\begin{equation}
		\begin{aligned}
			&m_b^{\star}=\max_{\mathbf{A}^{(b)}\in\mathcal{A}_b,\,\mathrm{supp}(\mathbf{A}^{(b)})\subset\mathcal{E}_b}\!|\mathbf{A}^{(b)}|, \\
			&\hat{\mathbf{A}}^{(b)}=\arg\min_{|\mathbf{A}^{(b)}|=m_b^{\star}}\sum_{(p,q)\in\mathcal{E}_b}c_{p,q}^{(b)}\,a_{p,q}^{(b)}.
		\end{aligned}
		\label{eq:max_cardinality_min_cost_assignment}
	\end{equation}
	Prioritizing cardinality over cost reflects our preference for track continuity. \eqref{eq:max_cardinality_min_cost_assignment} is a balanced linear assignment solved by the Hungarian algorithm \cite{munkres1957algorithms,emami2018machine}.
	
	The solution $\hat{\mathbf{A}}^{(b)}$ in \eqref{eq:max_cardinality_min_cost_assignment} propagates the global label through the buffer sequence. The first valid buffer $b_1$ is initialized by $g_{b_1}(k)=k$ for $k=1,\dots,K_{b_1}$, and a running counter $G$ is set to $K_{b_1}$. For each subsequent valid buffer $b$, the label map $g_b$ is built from its predecessor as
	{\small
		\begin{equation}
			g_b(q)=
			\begin{cases}
				g_{b^{-}}(p), & \text{if } \hat{a}_{p,q}^{(b)}=1 \text{ for matched } p \;,\\[2pt]
				G+1,\;\; G\!\leftarrow\!G+1, & \text{otherwise}.
			\end{cases}
			\label{eq:label_propagation}
		\end{equation}
	}
	That is a matched current cluster inherits its predecessor's global label, whereas an unmatched current cluster is assigned a fresh global label and increments the counter. The aligned candidate label of symbol $i\in\mathcal{I}_b$ in buffer $b$ is then
	\begin{equation}
		\tilde{s}_i^{(b)}=g_b\bigl(\ell_i^{(b)}\bigr).
		\label{eq:aligned_candidate_label}
	\end{equation}
	A special treatment is given for fusing labels in the buffers' overlap regions.
	
	\subsubsection{Overlap Fusion}
	
	Overlap fusion combines the candidate labels of a symbol covered by several buffers into one. Let $\mathcal{B}(i)=\{b:i\in\mathcal{I}_b\}$, $|\mathcal{B}(i)|\ge 1$. Each $b\in\mathcal{B}(i)$ supplies a candidate label $\tilde{s}_i^{(b)}$ and a reliability weight $\omega_i^{(b)}>0$ measuring how well-resolved its underlying cluster is. The final label is the weighted plurality vote
	\begin{equation}
		\begin{aligned}
			&\hat{s}_i=\arg\max_{g}V_i(g), \\
			&V_i(g)=\sum_{b\in\mathcal{B}(i)}\bar{\omega}_i^{(b)}\,\mathbf{1}\bigl(\tilde{s}_i^{(b)}=g\bigr), \\
			&\bar{\omega}_i^{(b)}=\frac{\omega_i^{(b)}}{\sum_{b'\in\mathcal{B}(i)}\omega_i^{(b')}}.
		\end{aligned}
		\label{eq:overlap_fusion_vote}
	\end{equation}
	Let $k=\ell_i^{(b)}$. The cluster-center estimate is a Gaussian belief on the source location with diagonal covariance
	\begin{equation}
		\mathbf{V}_{b,k}=\mathrm{diag}\bigl(v_{\theta,b,k},\,v_{\delta,b,k}\bigr).
		\label{eq:center_covariance}
	\end{equation}
	The weight of buffer $b$ is taken proportional to the inverse volume of its confidence ellipse, equivalently the geometric-mean precision of the belief:
	\begin{equation}
		\omega_i^{(b)}=\bigl|\mathbf{V}_{b,k}\bigr|^{-1/2}=\frac{1}{\sqrt{v_{\theta,b,k}\,v_{\delta,b,k}}}.
		\label{eq:fusion_weight}
	\end{equation}
	The weight depends on symbol $i$ only through its host cluster $k=\ell_i^{(b)}$, so all symbols assigned to the same cluster in buffer $b$ share the same $\omega_i^{(b)}$. A buffer with a tight, well-resolved cluster contributes a large vote, while a buffer with a sparse or diffuse cluster contributes a small one. The vote in \eqref{eq:overlap_fusion_vote} is computed directly. Ties are broken first by the unweighted coverage count $C_i(g)=\sum_{b} \mathbf{1}(\tilde{s}_i^{(b)} = g)$, and then by the buffer whose center is closest to $t_i$, ensuring a deterministic output. The final global source label sequence is
	\begin{equation}
		\hat{\mathbf{S}}_{1:N}=\{\hat{s}_i\}_{i=1}^N.
	\end{equation}
	
	\subsubsection{Replacement rule for TDoA $+$ Slant Delay}
	The pipeline is applied with three local modifications. First, the per-component density in \eqref{eq:local_mixture_density} is replaced by a single bivariate Gaussian with a full $2 \times 2$ covariance matrix, resulting in $5$ distribution parameters per component. Consequently, the BIC penalty becomes $p(J) = 6J - 1$, the minimum support becomes $N_b \ge 6J$ and the order bound of \eqref{eq:adaptive_kmax} becomes
	\begin{equation}
		K_{\max,b}=\min(K_\mathrm{max}^{(g)},\lfloor N_b/6\rfloor).
	\end{equation}
	Second, the signature in \eqref{eq:cluster_signature} becomes $\mathbf{h}_{b,k}=(n_{b,k},\bar{\mathbf{z}}_{b,k},\mathbf{S}_{b,k})$ with sample covariance $\mathbf{S}_{b,k}\in\mathbb{R}^{2\times 2}$ and cluster-center covariance $\mathbf{V}_{b,k}=\mathbf{S}_{b,k}/n_{b,k}$. Third, the residual likelihood becomes a bivariate Gaussian. As a result, $Q_{p,q}^{(b)}$ is exactly $\chi^{2}_{2}$, and the fusion weight $\omega_i^{(b)}=|\mathbf{V}_{b,k}|^{-1/2}$ must not be factorized as $1/\sqrt{v_\tau v_\delta}$ because the off-diagonal entries of $\mathbf{V}_{b,k}$ are generally nonzero. We now present the specific features we use for clustering.
	
	\subsection{Feature Extraction}
	\label{sec:feature_extraction}
	For each detected whistle symbol $i$, the detector returns a time support $\mathcal{W}_i=[t_i^\mathrm{s},t_i^\mathrm{e}]$ and an effective frequency band $\mathcal{F}_i$.
	
	\subsubsection{Bearing Estimation}
	
	Let $\mathbf{r}_0$ be the receiver location and $\mathbf{q}_i$ the true source location at symbol $i$. Under the horizontal far-field approximation, the geometric bearing is
	\begin{equation}
		\begin{aligned}
			&\theta_i^{\,\mathrm{geo}}=\mathrm{atan2}\bigl(-u_{i,y},\,-u_{i,x}\bigr), \\
			&\mathbf{u}_i=\Pi_h(\mathbf{r}_0-\mathbf{q}_i)/\|\Pi_h(\mathbf{r}_0-\mathbf{q}_i)\|,
		\end{aligned}
		\label{eq:true_bearing_geometry}
	\end{equation}
	where $\Pi_h(\cdot)$ retains the two horizontal components. The estimation problem is to recover $\widehat{\theta}_i$ from the bandpass array signals on $\mathcal{W}_i$ and to wrap it into $[-\pi,\pi)$ in the global frame,
	\begin{equation}
		\theta_i=\mathrm{wrap}_{[-\pi,\pi)}\!\bigl(\widehat{\theta}_i+\theta_\mathrm{cal}(t_i)\bigr),
		\label{eq:bearing_calibration}
	\end{equation}
	where $\theta_\mathrm{cal}(t_i)$ is a slowly varying calibration term absorbing installation orientation, compass yaw, and the array-to-source sign convention.
	
	When a vector hydrophone is available, let $p_i(t)$ denote the bandpass pressure channel and $v_{x,i}(t), v_{y,i}(t)$ the two orthogonal horizontal particle velocity channels, with respective analytic signals $P_i(t), V_{x,i}(t), V_{y,i}(t)$. The acoustic intensity vector points from the receiver toward the source, and its angle provides the bearing estimate in the array frame:
	\begin{equation}
		\begin{aligned}
			&\widehat{I}_{*,i}=\frac{1}{|\mathcal{W}_i|}\int_{\mathcal{W}_i}\mathrm{Re}\!\bigl\{P_i(t)V_{*,i}^*(t)\bigr\}\,dt,\quad *\in\{x,y\}, \\
			&\widehat{\theta}_i=\mathrm{atan2}\bigl(\widehat{I}_{y,i},\,\widehat{I}_{x,i}\bigr),
		\end{aligned}
		\label{eq:vector_hydrophone_bearing}
	\end{equation}
	with any channel polarity or sign convention offset absorbed into $\theta_\mathrm{cal}$.
	
	For a rigid array with at least four elements at known positions $\{\mathbf{r}_m\}_{m=1}^M$ recording band-pass signals $\{y_m(t)\}_{m=1}^M$, a plane wave from direction $\theta$ induces relative delays $\Delta t_m(\theta)=(\mathbf{r}_m-\mathbf{r}_1)^{\!\top}\mathbf{a}(\theta)/c$, with steering vector $\mathbf{a}(\theta)=[\cos\theta,\sin\theta]^{\!\top}$ and $c$ the sound speed, recall assumed to be fixed. The bearing is the direction at which the pair-wise time-aligned correlations are jointly maximized,
	\begin{equation}
		\widehat{\theta}_i=\arg\max_{\theta\in[-\pi,\pi)}\sum_{m<n}\int_{\mathcal{W}_i}y_m(t)\,y_n\!\bigl(t+\Delta t_n(\theta)-\Delta t_m(\theta)\bigr)\,dt.
		\label{eq:array_bearing_estimation}
	\end{equation}

	\subsubsection{Time Difference of Arrival (TDoA) Estimation}
	
	For a two-element array with hydrophone positions $\mathbf{r}_1,\mathbf{r}_2$, baseline $\mathbf{d}=\mathbf{r}_2-\mathbf{r}_1$ and $D=\|\mathbf{d}\|$, the geometric TDoA of symbol $i$ with direct arrival direction $\mathbf{u}_i$ is
	\begin{equation}
		\begin{aligned}
			&\tau_i^{\,\mathrm{geo}}=\mathbf{d}^{\!\top}\mathbf{u}_i/c, \\
			&|\tau_i^{\,\mathrm{geo}}|\le D/c,
		\end{aligned}
		\label{eq:ideal_tdoa}
	\end{equation}
	which is a one dimensional projection of the direction vector onto the baseline. The estimation problem is to recover $\widehat{\tau}_i$ from the two channels of symbol $i$ over $\mathcal{W}_i$ and $\mathcal{F}_i$.
	
	Let $y_{1,i}(t),y_{2,i}(t)$ be the bandpass segments of symbol $i$ on the two channels. The cross-correlation function quantifies the alignment between the two channels as a function of the trial lag $\tau$,
	\begin{equation}
		\begin{aligned}
			&R_i(\tau)=\int_{\mathcal{W}_i}y_{1,i}(t)\,y_{2,i}(t+\tau)\,dt, \\
			&\widehat{\tau}_i=\arg\max_{\tau\in[-D/c,D/c]}R_i(\tau).
		\end{aligned}
		\label{eq:tdoa_xcorr}
	\end{equation}
	For a direct arrival impinging from $\mathbf{u}_i$, the channel-$2$ waveform is a time-shifted replica of channel-$1$ by $\tau_i^{\,\mathrm{geo}}$, so $R_i(\tau)$ attains its global maximum at $\tau=\tau_i^{\,\mathrm{geo}}$ within the geometric admissible interval $[-D/c,D/c]$. The slant delay feature is introduced next, which encodes a propagation geometry projection.

	\subsubsection{The Slant Delay Estimation}
	
	The slant delay is the arrival-time gap between the direct path and the first resolvable reflection path of the same symbol,
	\begin{equation}
		\delta_i=t_{i,1}-t_{i,0}>0,
		\label{eq:drtd_definition}
	\end{equation}
	where $t_{i,0}$ and $t_{i,1}$ are the unknown true direct path and first reflection arrival times. Under a constant sound speed approximation with sound speed $c$, $\delta_i=(L_{i,1}-L_{i,0})/c$ with $L_{i,0},L_{i,1}$ the direct and first reflection-path lengths, so $\delta_i$ is set by the source--receiver range and depth geometry. The estimation problem is to recover $\widehat{\delta}_i=\widehat{t}_{i,1}-\widehat{t}_{i,0}$ from the bandpass pressure segment $p_i(t)$ on $\mathcal{W}_i$.
	
	We use a simple energy detector. The smoothed analytic envelope of $p_i$ is
	\begin{equation}
		\begin{aligned}
			&a_i(t)=\bigl|p_i(t)+\mathcal{H}\{p_i(t)\}\bigr|, \\
			&\bar{a}_i(t)=(a_i*h_s)(t),
		\end{aligned}
		\label{eq:analytic_envelope}
	\end{equation}
	where $\mathcal{H}\{\cdot\}$ is the complex Hilbert transform and $h_s$ is a nonnegative smoothing kernel with unit integral. Both arrivals are selected from the set of local peaks of $\bar{a}_i$, that is, the sample times at which the envelope is strictly greater than its immediate temporal neighbors, so that the selected instants correspond to actual energy maxima rather than points on a rising or falling slope. The direct-path arrival $\widehat{t}_{i,0}$ is the earliest such local peak in $\mathcal{W}_i$ exceeding $\gamma_0\!\cdot\!\max_{u\in\mathcal{W}_i}\bar{a}_i(u)$. The first reflection $\widehat{t}_{i,1}$ is the strongest local peak in the post-direct search window $\mathcal{R}_i=[\widehat{t}_{i,0}+T_g,\,\widehat{t}_{i,0}+T_\mathrm{max}^\delta]$ that exceeds a relative threshold:
	\begin{equation}
		\begin{aligned}
			&\widehat{t}_{i,1} = \arg\max_{t\in\mathcal{R}_i}\;\bar{a}_i(t) \\
			&\text{s.t.} \quad \bar{a}_i(t)\ge\gamma_1\,\bar{a}_i(\widehat{t}_{i,0}) .
		\end{aligned}
		\label{eq:reflection_peak_detection}
	\end{equation}
	Here, $T_g$ is a guard interval that prevents the direct-peak skirt from being selected and, by trial and error, we choose the value of $0.15$~ms, $T_\mathrm{max}^\delta \approx 2z_r/c$ is the maximum physically admissible slant delay for a surface bounce at receiver depth $z_r$, and $\gamma_0,\gamma_1 \in (0,1)$ are amplitude thresholds for which, by trial and error, we choose values 0.08 and $0.25$, respectively. The slant delay estimate is $\widehat{\delta}_i=\widehat{t}_{i,1}-\widehat{t}_{i,0}$.
	As shown in our results, the bearing or TDoA features and the slant delay feature complement each other in distinguishing between dolphin sources.
	
	\subsection{Complexity Analysis}
	\label{sec:complexity}
	
	The overall cost of the algorithm is dominated by the per-buffer EM clustering, while the cross-buffer matching and overlap fusion contribute lower-order terms. Within buffer $b$, one EM iteration costs $\mathcal{O}(N_b J)$ for $J$ candidate components. The BIC sweep runs for $J = 1,\dots, K_{\max,b}$ with $K_{\max,b} \le \lfloor N_b / 5 \rfloor$, so the worst-case cost per buffer is $\mathcal{O}(N_b K_{\max,b}^{2}\, T_\mathrm{EM} R_\mathrm{init})$, where $T_\mathrm{EM}$ is the average number of iterations to convergence and $R_\mathrm{init}$ is the number of random initializations. The Hungarian matching contributes $\mathcal{O}(L_b^{3})$ with $L_b \le K_\mathrm{max}^{(g)}$, and the overlap fusion is $\mathcal{O}(N_b)$. With $B=\mathcal{O}(N / N_b)$ buffers in a recording of $N$ symbols, the total worst-case complexity is
	\begin{equation}
		\mathcal{O}\!\Bigl(\textstyle\sum_{b=1}^{B}N_b\,K_{\max,b}^{2}\,T_\mathrm{EM}R_\mathrm{init}\;+\;B\,(K_\mathrm{max}^{(g)})^{3}\;+\;{N}\Bigr).
		\label{eq:total_complexity}
	\end{equation}
	For a fixed buffer length $T_\mathrm{buf}$, both $N_b$ and $K_{\max,b}$ are bounded by problem specific constants, so \eqref{eq:total_complexity} grows linearly with $N$.
	\begin{figure}
		\centering
		\includegraphics[width=0.4\columnwidth]{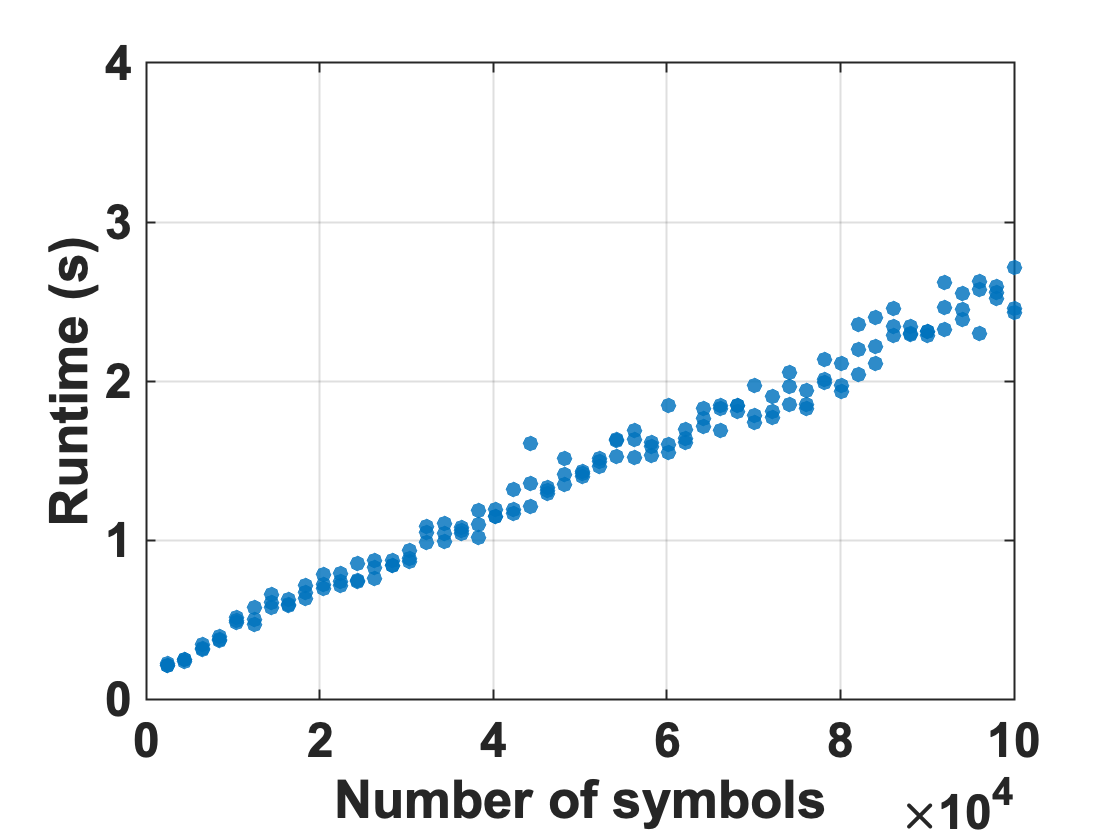}
		\caption{Runtime measurement as a function of the number of whistles per $600$~s frame. The runtime scales approximately linearly, requiring about 3 s for $10^{5}$ symbols per frame.}
		\label{fig:runtime}
	\end{figure}
	
	We confirm this near-linear scaling empirically. Fig.~\ref{fig:runtime} shows the measured run time of the proposed algorithm on $600$~s recording frames, with the number of whistle symbols per frame varied from $2\!\times\! 10^{3}$ to $10^{5}$, executed in MATLAB on a 2021 MacBook Pro with an Apple M1 Pro processor and 16 GB RAM. The run time scales approximately linearly, requiring about 3~s for $10^{5}$ symbols per frame; the small dispersion in the upper-right region reflects the BIC sweep occasionally reaching higher candidate orders in densely populated buffers. The polynomial complexity and the demonstrated short run time make the proposed approach suitable for segmenting long recordings and for continuous monitoring.
	
	\section{Results}
	\label{sec:results}
	In this section, we examine the performance of our solutions in simulations, controlled sea experiments, and real dolphin recordings. As a performance metric, we use the macro-F1 score, calculated by computing the F1 score for each source and averaging across sources, which quantifies the precision-recall trade-off. Multiple Object Tracking Accuracy (MOTA) was also calculated and showed the same trend as the macro-F1 score. To the best of our knowledge, there is no previous solution for the unsupervised clustering of dolphin whistle sources. As a baseline, we consider two clustering approaches. The first is K-means clustering, which is suitable for a feature space with distinct centroids. The second is expectation maximization (EM) clustering, which is appropriate when the features form distinct distributions. Since both our method and the reference approaches require a specified number of clusters, we use the known number of dolphins as an upper bound. This prior can be avoided by running the clustering solution for several assumed cluster numbers and selecting the one with the best convergence. We report that, using these strategies, only a negligible decrease in performance was observed, but at the cost of increased complexity.
	
	\subsection{Simulation}
	
	\subsubsection{Simulation Setup}
	\begin{table*}
		\centering
		\caption{The five motion scenarios used in (\ref{eq:rw}); for the mixed (MX) scenario, the $N_D$ dolphins are cyclically assigned to one of the four basic profiles. Here low-turning or high-turning refers to the per-emission turning rate $\sigma_\psi$. Since the heading follows a random walk, even low-turning tracks accumulate an appreciable heading change ($\approx\sigma_\psi\sqrt{N}$) over a long run.}
		\label{tab:sim}
		\setlength{\tabcolsep}{0pt}
		\begin{tabular}{l c c c}
			% \midrule
			% Bathymetry & Emissions per dolphin & $\Delta t$ (s) & Observation duration (s) \\
			% \midrule
			% San Diego  & 120 & 0.5 & 60  \\
			% Taiwan     & 120 & 0.5 & 60 \\
			\midrule
			Scenario      & Speed (m/s) & $\sigma_\psi$ (deg) & $\sigma_z$ (m) \\
			\midrule
			slow speed with low-turning (SL) & $[2.0,\,3.5]$ & $3$  & $0.20$ \\
			slow speed with high-turning (SH)  & $[2.0,\,3.5]$ & $22$ & $0.25$ \\
			fast speed with low-turning (FL) & $[4.0,\,6.5]$ & $3$  & $0.22$ \\
			fast speed with high-turning (FH)  & $[4.0,\,6.5]$ & $22$ & $0.28$ \\
			mixed (MX)         & \multicolumn{3}{c}{cyclic assignment to the 4 profiles above} \\
			\bottomrule
		\end{tabular}
	\end{table*}
	
    The simulated dolphins emit whistles in a prediction fashion (see parameters in Table~\ref{tab:sim}) and the signals propagate to the receiver using the Bellhop propagation model. The randomized location of the dolphins make sure to include overlaps in the created dataset. 
    Two different bathymetric profiles are considered. The San Diego map is diverse, and the resulting slant delay exhibits a complex spatial structure compared to the flatter Taiwan map, which represents the eastern coast of Taiwan. The water depth in the selected area of the Taiwan map ranges from approximately $4,530$~m to $4,750$~m, capturing a descending deep sea basin floor.  Comparing the performance of the clustering solution in both environments allows us to test whether the method is robust to different propagation regimes.
	
	We simulate $N_D\in\{2,\dots,9\}$ dolphins moving by a discrete random-walk (RW) model. The position of dolphin $m$ at its $n$-th emission updates as
	\begin{equation}
		\begin{aligned}
			&\mathbf{q}_{m,n+1}=\mathbf{q}_{m,n} + v_{m}\,\Delta t\,[\cos\psi_{m,n},\,\sin\psi_{m,n},\,0]^\top + \boldsymbol{\epsilon}_{m,n}, \\
			&\psi_{m,n+1}=\psi_{m,n}+\eta_{m,n},\qquad \eta_{m,n}\sim\mathcal{N}(0,\sigma_\psi^2),
		\end{aligned}
		\label{eq:rw}
	\end{equation}
	where $\Delta t$ is the emission interval, and $v_{m}$ and $\psi_{m,n}$ are the source moving speed and heading, with depth perturbation $\boldsymbol{\epsilon}_{m,n} = [0, 0,\zeta_{m,n}]^\top$, respectively. For each dolphin, the speed is randomized per random walk by $v_{m} \sim \mathcal{U}(v_\text{low}, v_\text{high})$~ [m/s], and at each emission, the heading increment is drawn as $\eta_{m,n} \sim \mathcal{N}(0,\sigma_\psi^2)$~ [deg], and the depth perturbation as $\zeta_{m,n} \sim \mathcal{N}(0,\sigma_z^2)$~ [m]. The triplet $([v_\text{low}, v_\text{high}],\sigma_\psi,\sigma_z)$ defines a motion scenario. Table~\ref{tab:sim} lists the parameter values for five scenarios that cover slow and fast swimming combined with weak or strong turning, along with a mixed scenario in which the $N_D$ dolphins are cyclically assigned to one of the four basic profiles. The bearing is computed from the source–receiver horizontal geometry, and the slant delay is computed by applying the widely used Bellhop propagation model to calculate the delay between the direct path and the first dominant reflection. Gaussian noise is added to the bearing and slant delay measurements. The two features are perturbed jointly according to the schedule in Table~\ref{tab:noise}, in which $\sigma_\theta$ [deg] and $\sigma_{\mathrm{slant}}$ [ms] increase together from level~1 to level~9. Next, we explore the simulation results in terms of sensitivities to overlap, propagation environment, noise level, number of source dolphins, and motion type.
	\begin{table}
		\centering
		\caption{Joint feature perturbation schedule. At each level, zero-mean Gaussian noise with standard deviation $\sigma_\theta$ is added to the bearing and $\sigma_{\mathrm{slant}}$ to the slant delay. The same schedule is used for both bathymetries. Level~5 is the operating point used for the scenario and source number analyses.}
		\label{tab:noise}
		\begin{tabular}{c c c}
			\toprule
			Level & $\sigma_\theta$ (deg) & $\sigma_{\mathrm{slant}}$ (ms) \\
			\midrule
			1  & 0.5  & 0.005 \\
			2  & 1    & 0.01  \\
			3  & 3    & 0.02  \\
			4  & 5    & 0.05  \\
			5  & 8    & 0.10  \\
			6  & 12   & 0.20  \\
			7  & 18   & 0.35  \\
			8  & 25   & 0.50  \\
			9 & 35   & 0.75  \\
			\bottomrule
		\end{tabular}
	\end{table}
    \begin{figure}
		\centering
		\includegraphics[width=0.4\columnwidth]{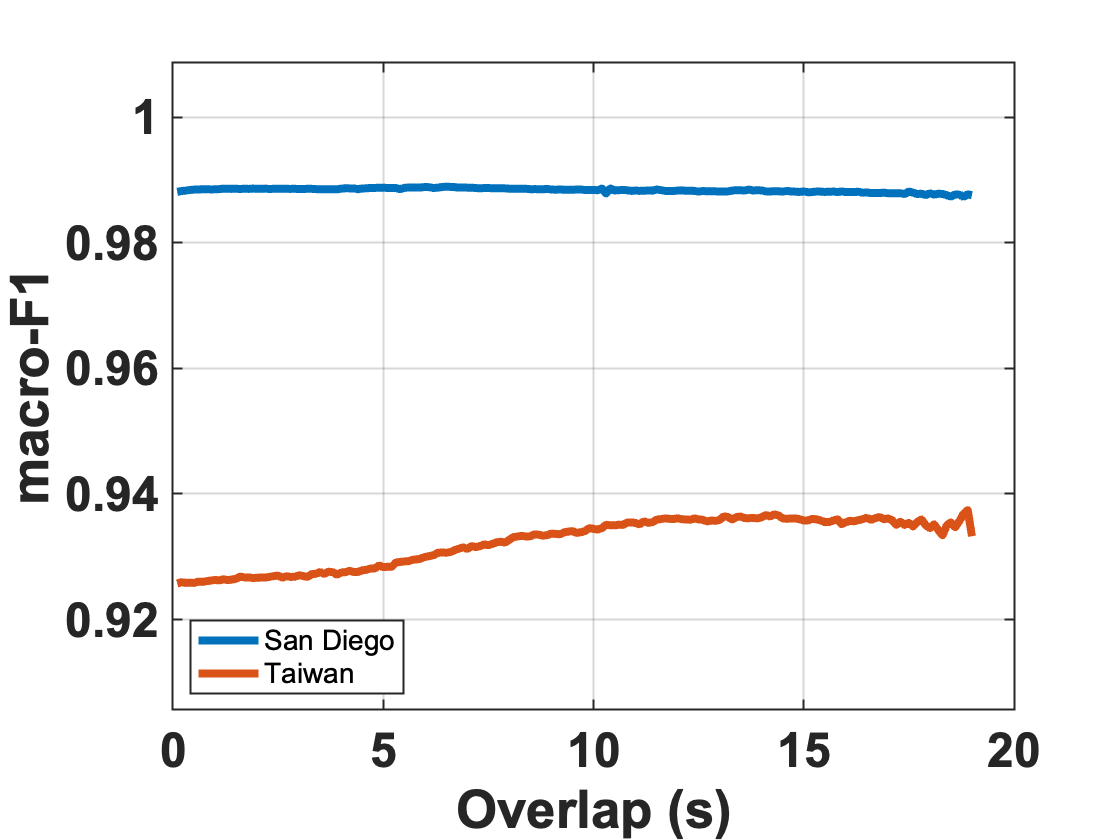}
		\caption{Macro-F1 versus buffer overlap for the two bathymetries. No noise is added, feature and the values are averaged over the number of dolphins and motion scenarios.}
		\label{fig:simulation_combined_overlap_sensitivity}
	\end{figure}
	
	\subsubsection{Simulation Results}
	Throughout the simulation study, we follow a fixed evaluation protocol. We first vary the buffer overlap on features without noise, averaging each operating point over all source counts $N_D \in \{2,\dots, 9\}$ and the five motion scenarios in Table~\ref{tab:sim}. With the overlap fixed, we then vary the noise level, again averaging each point over all $N_D$ and motion scenarios. Finally, at fixed overlap and noise level, we report the per-scenario macro-F1.
	
	The sensitivity of the proposed method to the overlap parameter in the Taiwan and San Diego propagation environments is shown in Fig.~\ref{fig:simulation_combined_overlap_sensitivity}. In the San Diego environment, we observe stable local clusters and cross-buffer matches even at small overlap values. This robustness is less evident in the results using the bathymetry from Taiwan. There, the deep water geometry and the presence of sources with similar features make single-buffer decisions more error-prone. Increasing the overlap allows each symbol to be covered by several buffers, so the fusion vote suppresses occasional misassignments. In the following, we fix the selected overlap value at $3.2$~s for San Diego and $6.6$~s for Taiwan.
	
\begin{figure*}
		\centering
		\begin{subfigure}[b]{0.4\textwidth}
			\centering
			\includegraphics[width=\textwidth]{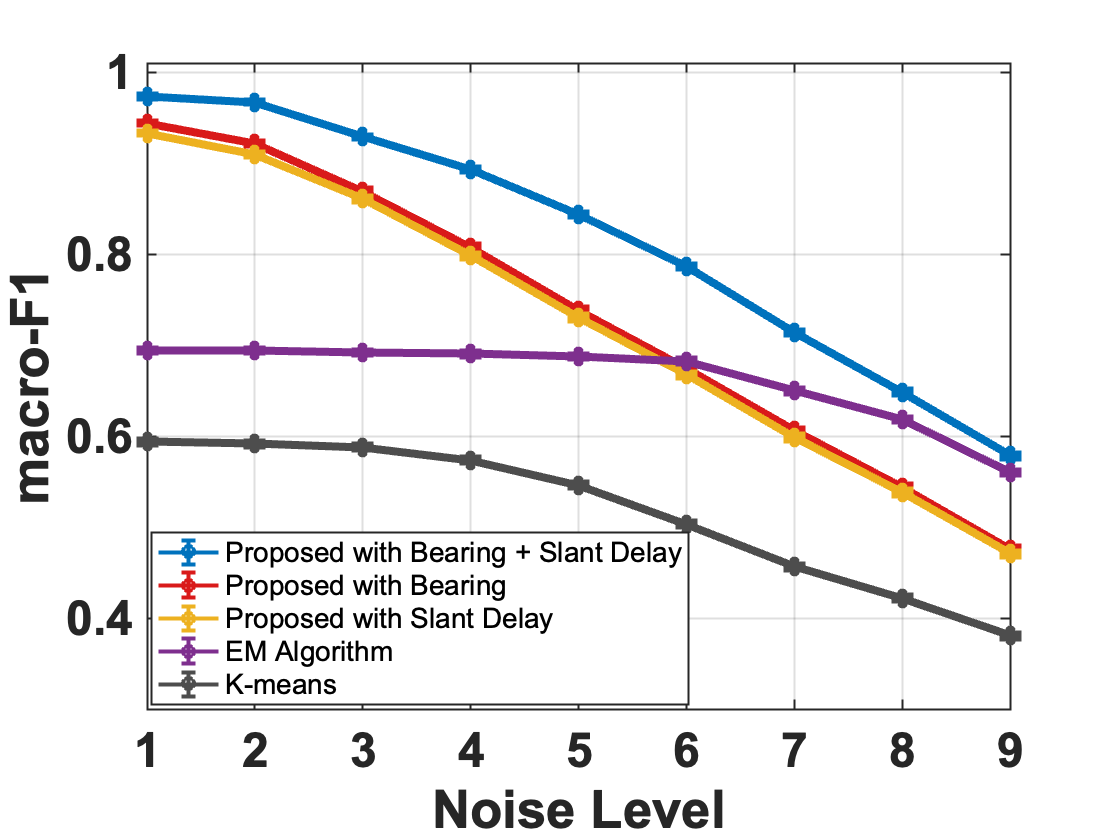}
			\caption{ }
			\label{fig:Simulation_SD_noise_sensitivity}
		\end{subfigure}
		\begin{subfigure}[b]{0.4\textwidth}
			\centering
			\includegraphics[width=\textwidth]{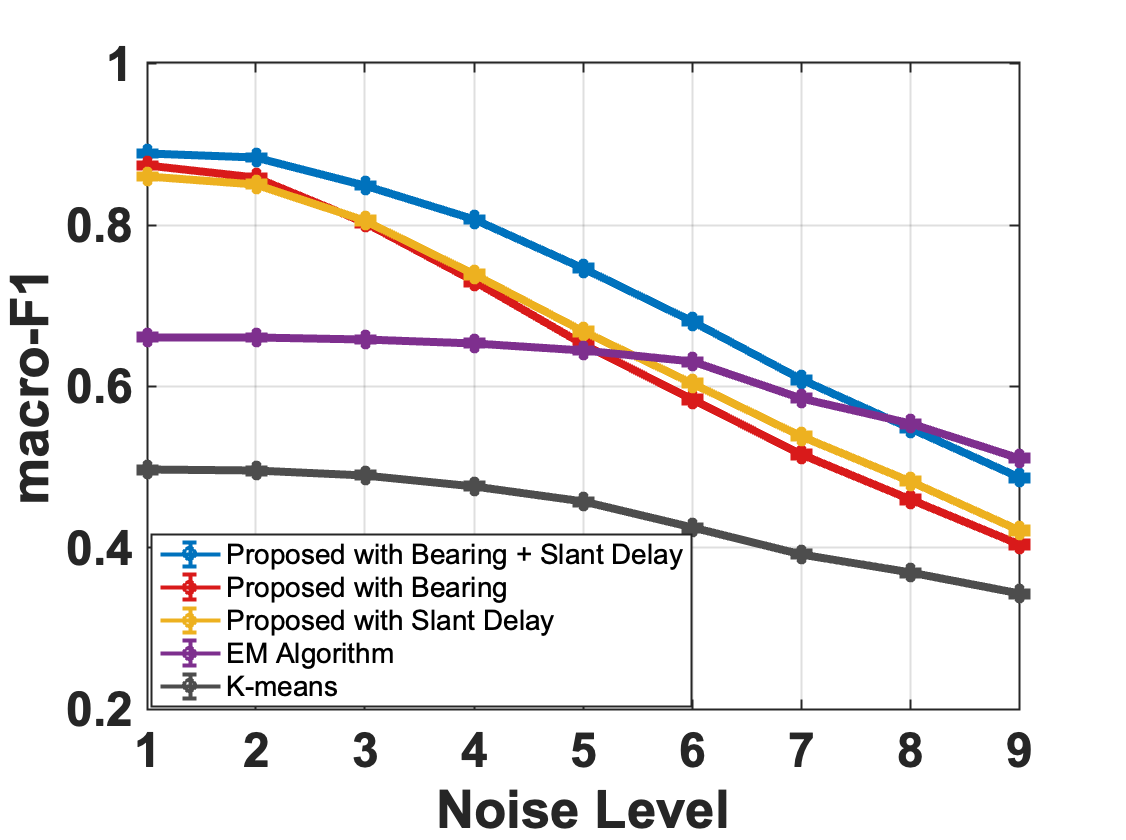}
			\caption{ }
			\label{fig:Simulation_TW_noise_sensitivity}
		\end{subfigure}
		\begin{subfigure}[b]{0.4\textwidth}
			\centering
			\includegraphics[width=\textwidth]{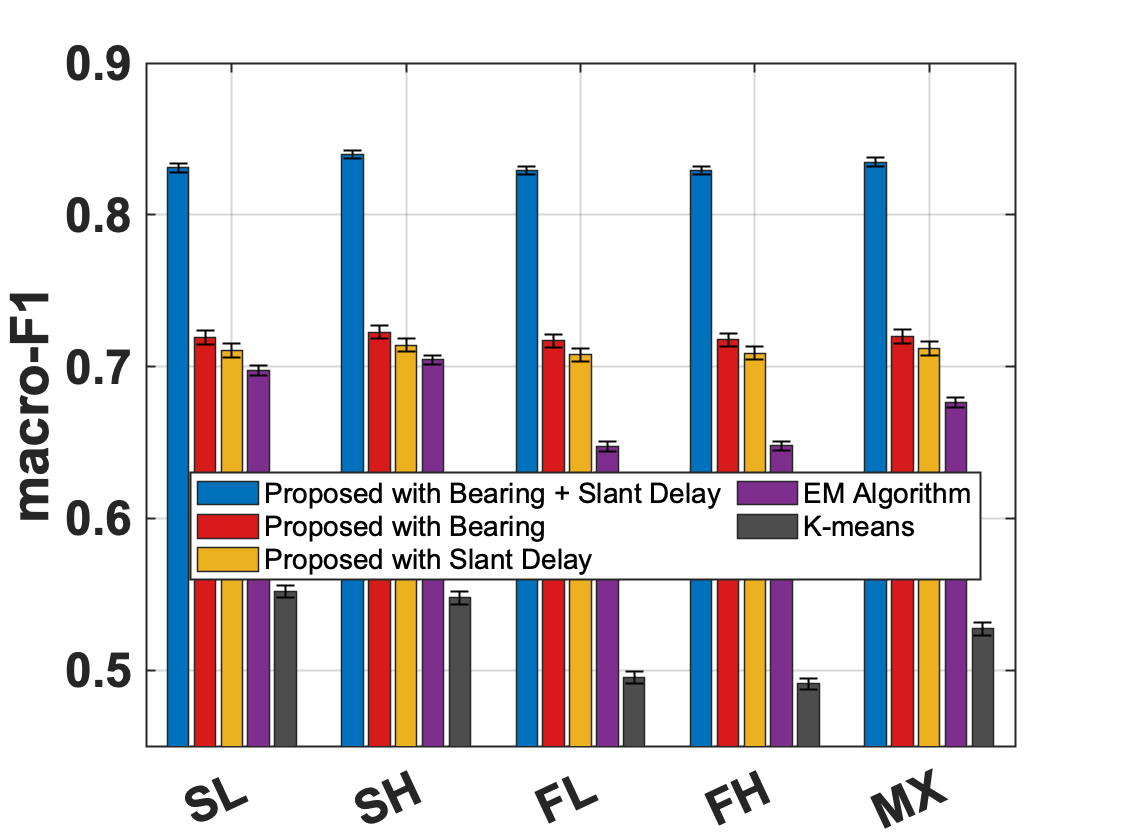}
			\caption{ }
			\label{fig:Simulation_SD_scenario_sensitivity}
		\end{subfigure}
		\begin{subfigure}[b]{0.4\textwidth}
			\centering
			\includegraphics[width=\textwidth]{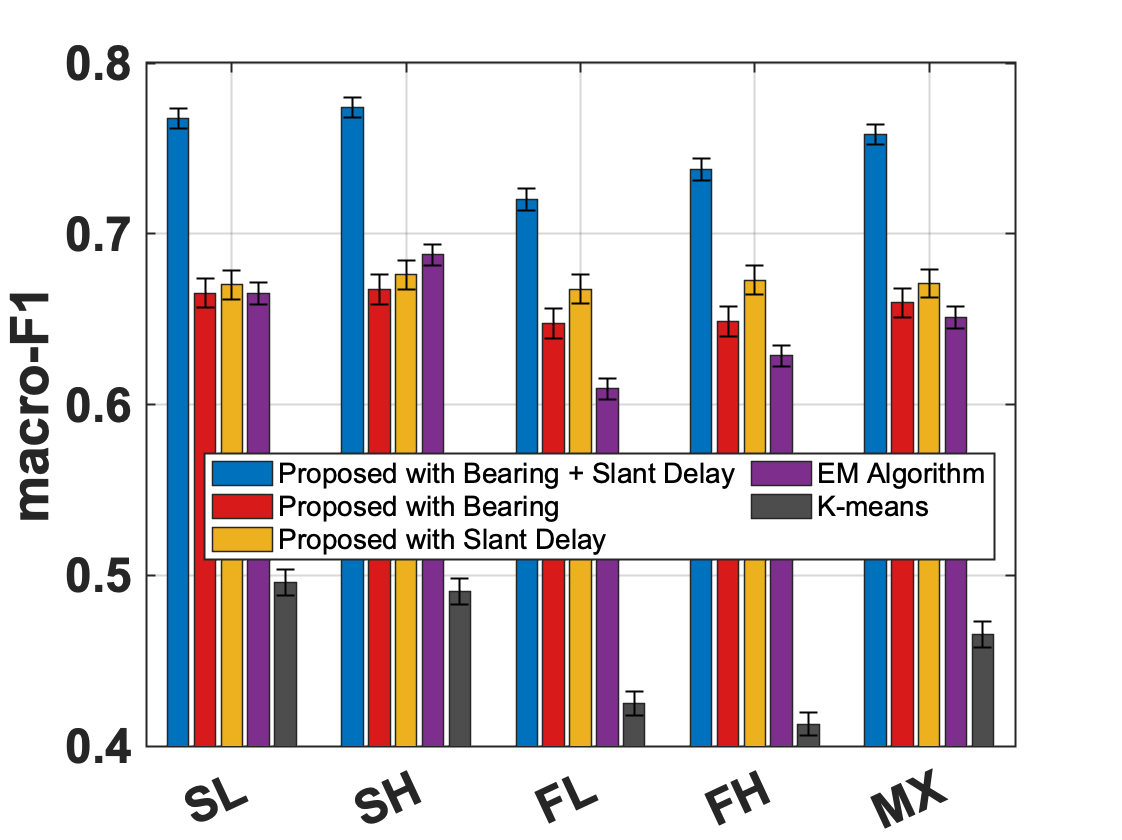}
			\caption{ }
			\label{fig:Simulation_TW_scenario_sensitivity}
		\end{subfigure}
		\caption{Macro-F1 scores for the five methods at a fixed overlap of 3.2 s for San Diego and 6.6 s for Taiwan. (a) and (b) is Macro-F1 versus noise level in Table~\ref{tab:noise}, averaged over all source numbers and motion scenarios; (c) and (d) is Macro-F1 across different motion scenarios, averaged over dolphin numbers from 2 to 9. Panels (a) and (c) represent San Diego, and (b) and (d) represent Taiwan. Error bars denote the 95\% confidence interval of the mean.}
		\label{fig:Simulation_noise_and_scenario_sensitivity}
	\end{figure*}

	The sensitivity of the proposed method, K-means, and EM clustering is evaluated by adding Gaussian noise to the selected features. The results are shown in Fig.~\ref{fig:Simulation_SD_noise_sensitivity} and Fig.~\ref{fig:Simulation_TW_noise_sensitivity} as a function of the noise standard deviation. In the San Diego case, the proposed method increases from about $0.58$ to about $0.98$ at low noise levels and consistently outperforms the reference schemes. When only bearing information (\textit{bearing only}) or only the slant delay feature (\textit{slant delay only}) is used, performance approaches that of the joint approach only at very low noise levels. Comparing the results for the two bathymetries, we observe that the one corresponding to Taiwan is more challenging overall, with lower macro-F1 values than San Diego. This is due to better separation of the channel's features in diverse sea environments. Acceptable performance is achieved in both environments for a bearing noise standard deviation of $\sigma_\theta=8^\circ$ and a slant delay noise standard deviation of $\sigma_{\mathrm{slant}}=0.1$~ ms. In addition to noise, the motion pattern may also affect performance.
	
	At a fixed overlap and added noise level of $\sigma_\theta = 8^\circ$, $\sigma_{\mathrm{slant}} = 0.1$~ ms, we compare the five method variants across different dolphin motion patterns. Results are shown in Fig.~\ref{fig:Simulation_SD_scenario_sensitivity} and Fig.~\ref{fig:Simulation_TW_scenario_sensitivity}. In the figure, \textit{SL} denotes slow speed with low turning, \textit{SH} denotes slow speed with high turning, \textit{FL} denotes fast speed with low turning, and \textit{FH} denotes fast speed with high turning. The figure demonstrates that the proposed joint features method achieves the highest F1 score under all motion patterns. The single feature variants and the EM approaches perform similarly, while K-means shows the lowest performance. This advantage only slightly decreases for the Taiwan dataset. The results using the less diverse Taiwan bathymetry show a preference for slower motion types, with fewer differences observed in the channel's features. Surprisingly, we do not observe such sensitivity for the more diverse San Diego dataset. This is likely because channel diversity in complex bathymetries is already high and thus less affected by different motion patterns. The influence of dolphin numbers on the performance of the proposed method is considered next.

    \begin{figure*}
		\centering
		\begin{subfigure}[b]{0.4\textwidth}
			\centering
			\includegraphics[width=\textwidth]{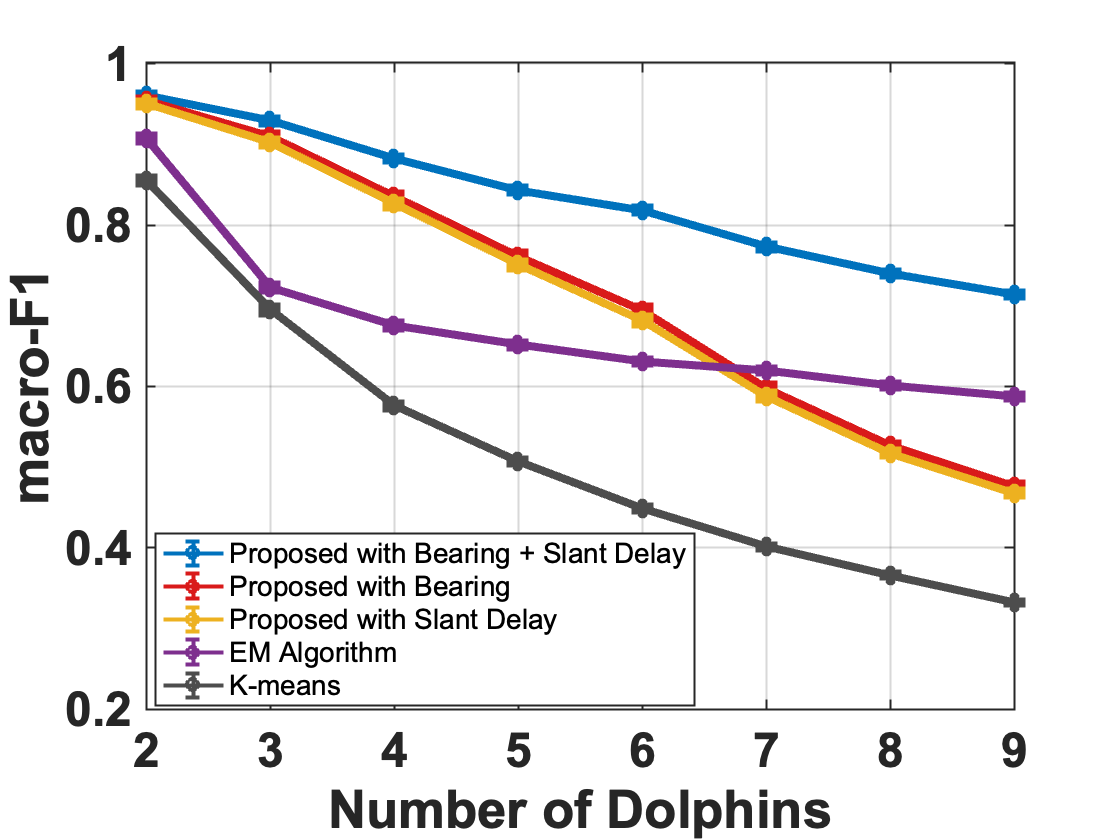}
			\caption{ }
			\label{fig:Simulation_SD_numbers_sensitivity}
		\end{subfigure}
		\begin{subfigure}[b]{0.4\textwidth}
			\centering
			\includegraphics[width=\textwidth]{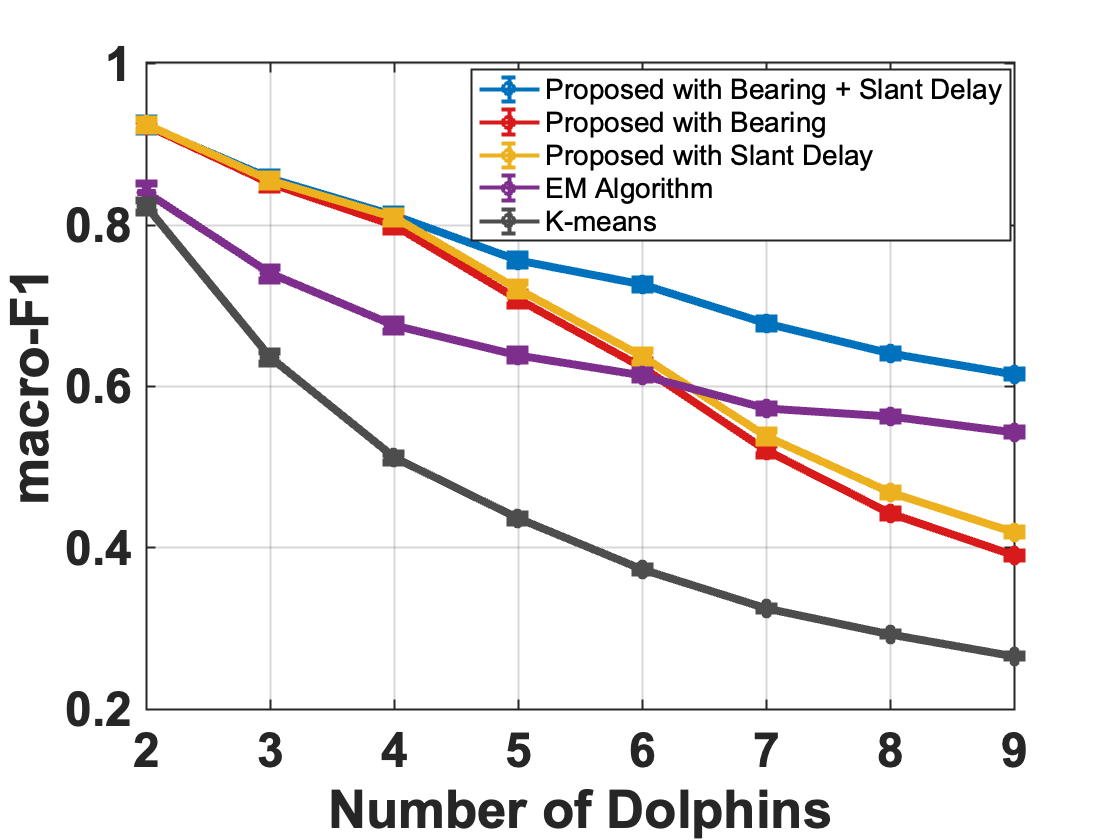}
			\caption{ }
			\label{fig:Simulation_TW_numbers_sensitivity}
		\end{subfigure}
		\caption{Macro-F1 versus number of sources for the five methods with fixed overlap of $3.2$~s and $6.6$~s. Each numbers of dolphin averaged over the five motion scenarios. (a) San Diego; (b) Taiwan. Error bars denote the $95\%$ confidence interval of the mean.}
		\label{fig:Simulation_numbers_sensitivity}
	\end{figure*}

	The F1 score as a function of the number of source dolphins is shown in Fig.~\ref{fig:Simulation_numbers_sensitivity}. Since the number of dolphins determines the number of clusters, performance decreases as the number of classes increases. Although the other four reference schemes show high sensitivity, the decrease in performance for the joint approach is not severe. This performance decrease is similar in both the San Diego and Taiwan environments. In the San Diego map, the joint and partial feature approaches yield similar results when the number of sources is two. In low-density conditions, a single feature is sufficient to provide a strong degree of distinction. As the number of sources increases, the single feature methods decline more rapidly and fall below the EM reference.
    
	\begin{figure*}
		\centering
		\begin{subfigure}[b]{0.4\textwidth}
			\centering
			\includegraphics[width=\textwidth]{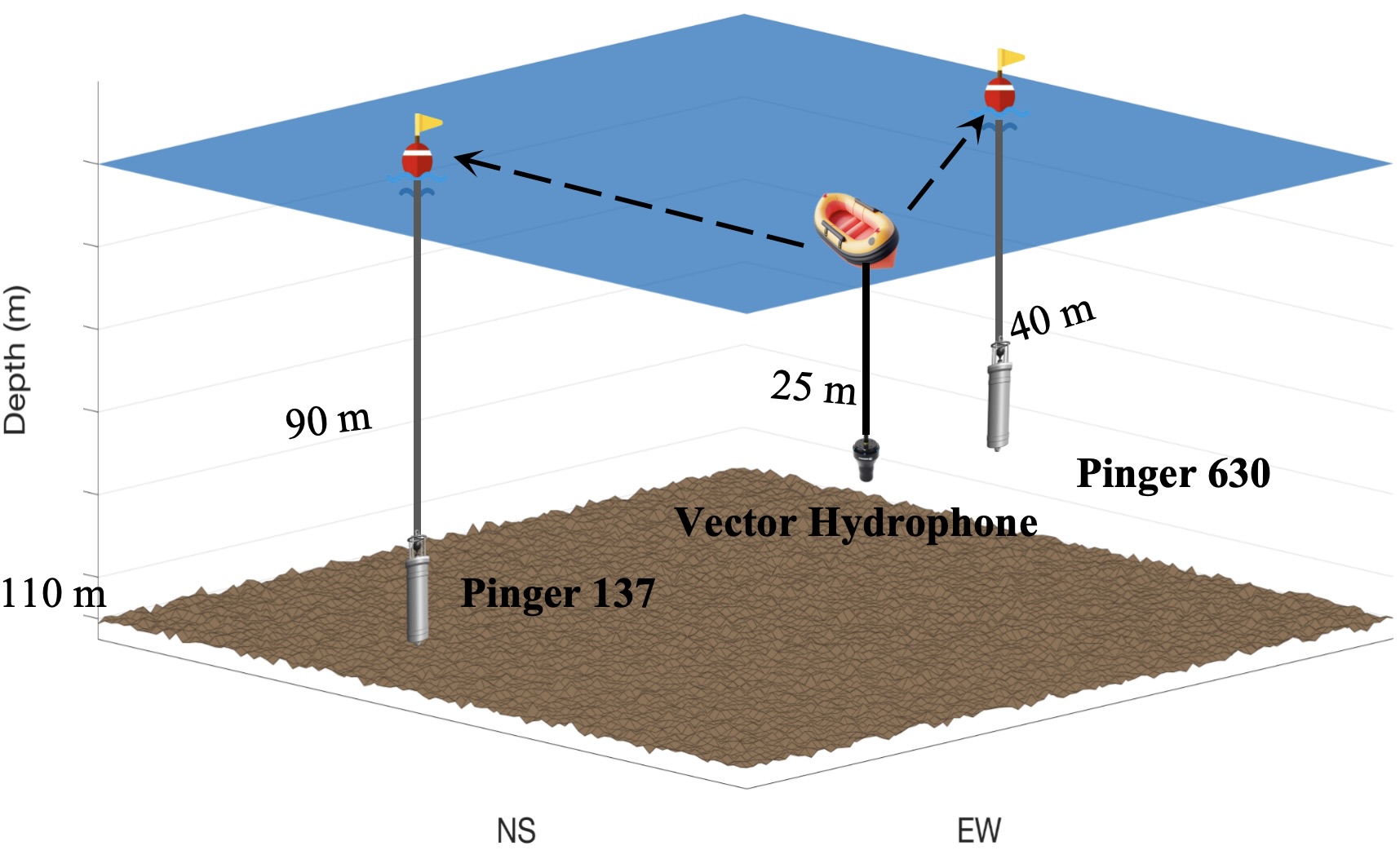}
			\caption{ }
			\label{fig:sea_exp_overview}
		\end{subfigure}
		\begin{subfigure}[b]{0.4\textwidth}
			\centering
			\includegraphics[width=\textwidth]{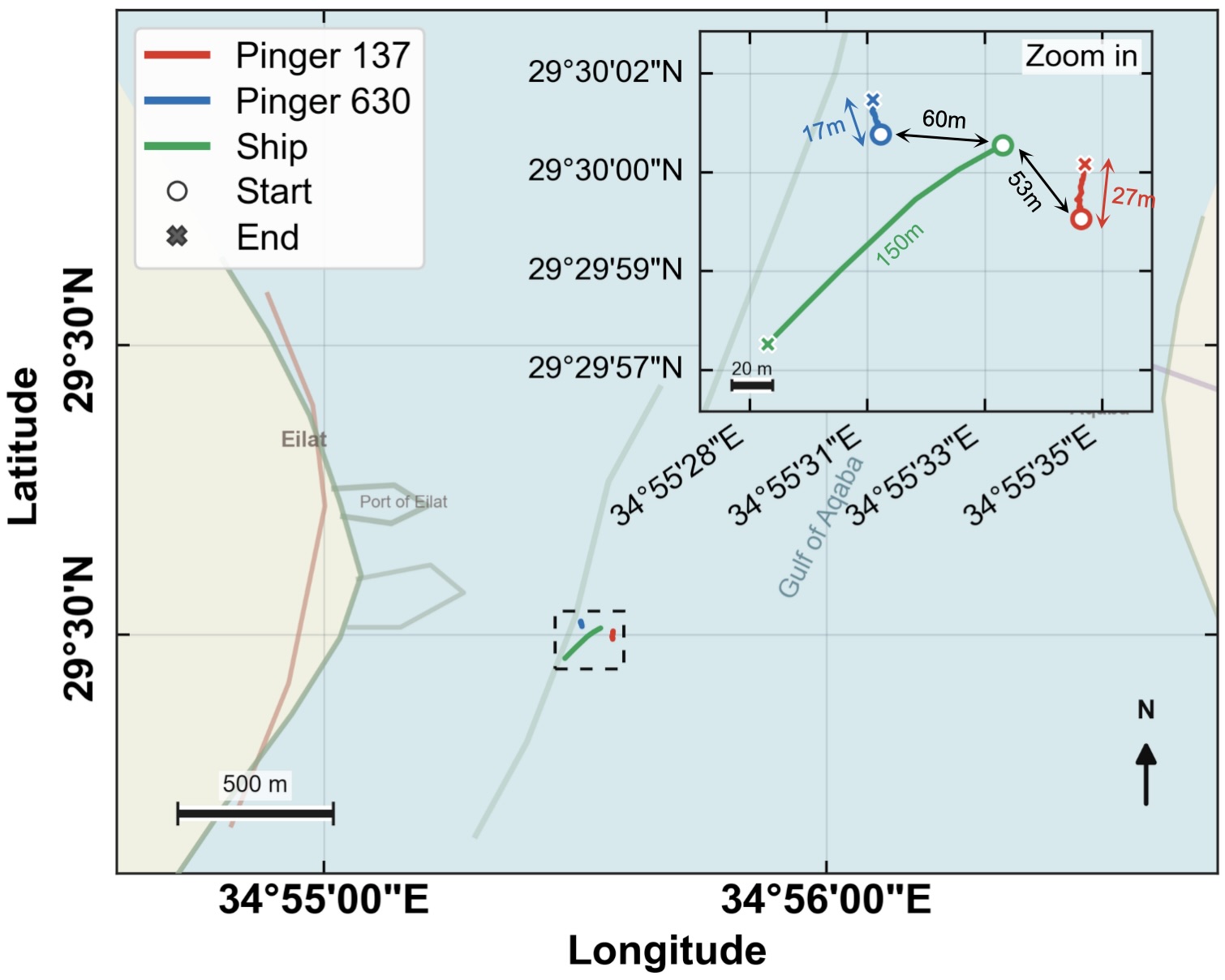}
			\caption{ }
			\label{fig:GPS_ill}
		\end{subfigure}
		\begin{subfigure}[b]{0.4\textwidth}
			\centering
			\includegraphics[width=\textwidth]{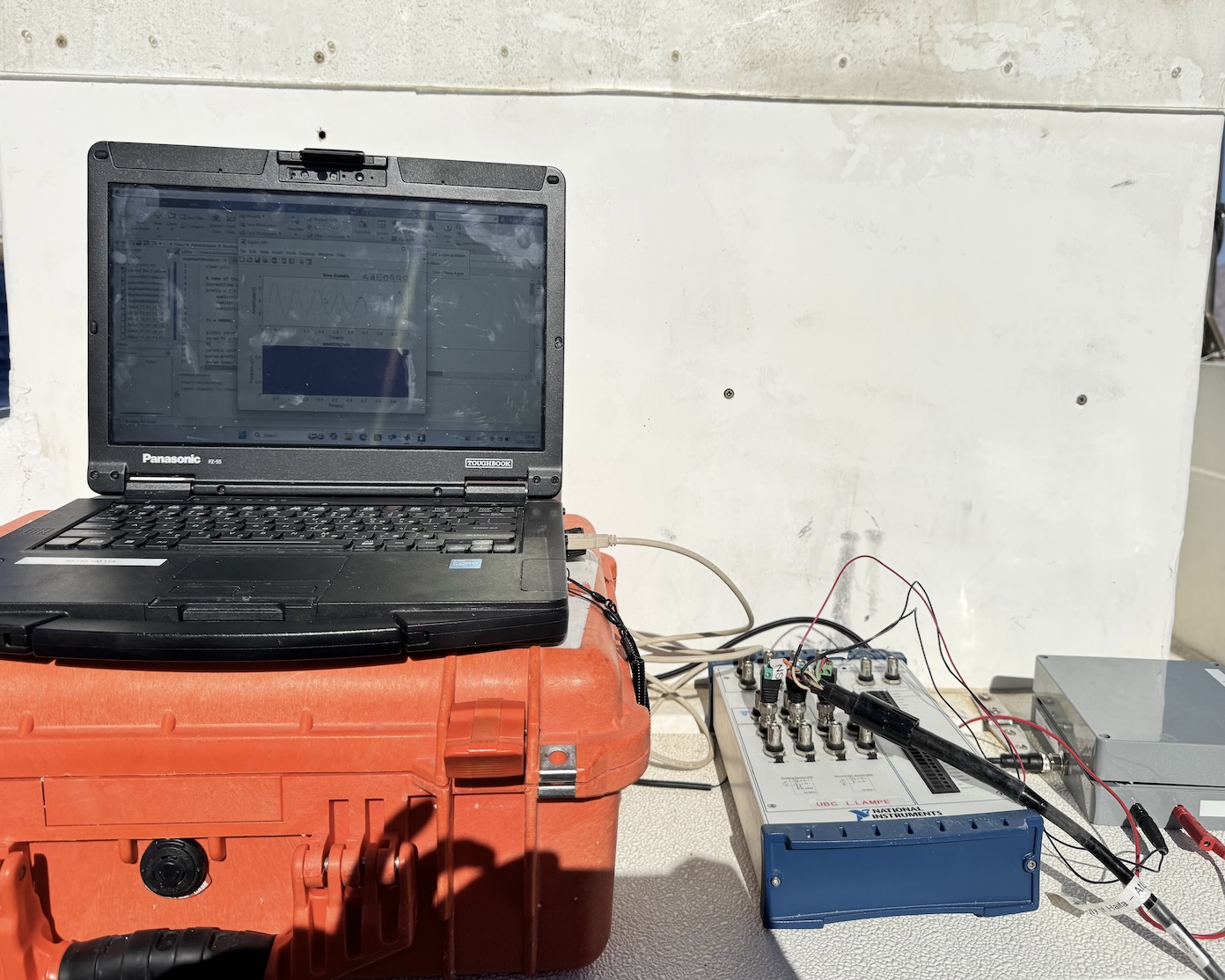}
			\caption{ }
			\label{fig:Recording_setup}
		\end{subfigure}
		\begin{subfigure}[b]{0.4\textwidth}
			\centering
			\includegraphics[width=\textwidth]{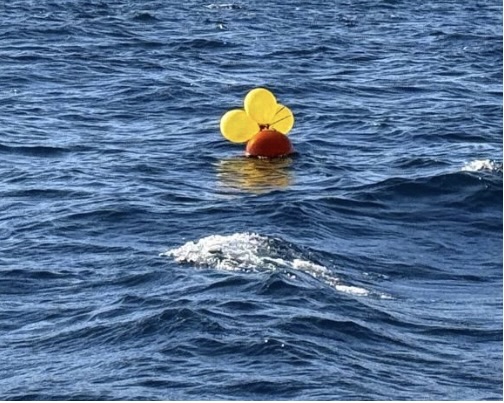}
			\caption{ }
			\label{fig:Pinger_in_sea}
		\end{subfigure}
		\caption{Setup of the controlled sea experiment, conducted off Eilat in the Red Sea at a water depth of about 110 m. Panel (a) is a schematic of the layout, with two narrowband pingers deployed at 40 m and 90 m and a vector-hydrophone receiver at 25 m; (b) shows the GPS trajectories of the two pingers and the recording ship, all of which drifted freely during the experiment; (c) is the recording desktop and acquisition hardware; and (d) is a pinger tethered beneath a surface float. Because the source identities are known, this experiment provides labeled ground truth for evaluating the separation under realistic conditions.}
		\label{fig:control_exp_setup}
	\end{figure*}
	
	\subsection{Controlled Sea Experiment}
	
	While our numerical simulation provided a statistical performance analysis, it is necessary to demonstrate the performance of our scheme in real sea conditions, where realistic bearing and slant delay errors occur. To this end, we conducted a controlled sea experiment in which the identity of the source of the emitted signals is known.
	
	\subsubsection{Experiment Setup}
    \begin{table*}
		\centering
		\caption{Controlled sea experiment setup.}
		\label{tab:ctrl}
		\begin{tabular}{ll}
			\toprule
			Parameter & Value \\
			\midrule
			Depth        & around $110$~m  \\
			Receiver            & vector hydrophone, depth $25$~m \\
			Pinger 630          & $13.000$~kHz, $25$~ms pulse, $0.5$~s interval, depth $40$~m \\
			Pinger 137          & $13.125$~kHz, $25$~ms pulse, $1.0$~s interval, depth $90$~m \\
			Sampling rate       & $40$~kHz \\
			\bottomrule
		\end{tabular}
	\end{table*}
	Our sea experiment included two acoustic pingers deployed at different depths and locations, each periodically emitting single carrier pulses. The experiment was conducted in Eilat, Red Sea, at a water depth of $110$~m. Two narrowband pingers (Model No. 1260 PINGER from Online Electronics Ltd., Aberdeen) were deployed at depths of $40$~m and $90$~m from buoys equipped with GPS loggers and emitted single carrier pulses periodically. Pinger 630 emitted $13.000$~kHz pulses at a pulse rate of $0.5$~s. Pinger 137 emitted $13.125$~kHz pulses at a pulse rate of $1.0$~s. A receiver was deployed from a boat roughly midway between the pingers at a depth of $25$~m. The receiver was a vector hydrophone with a hydrophone and two orthogonal particle-velocity sensors to calculate the bearing angle. Both the pingers and receivers drifted freely. Fig.~\ref{fig:control_exp_setup} shows the experimental layout, the GPS trajectories and pictures of the recorder and emitter setup. 
    The main parameters of the controlled experiment are listed in Table~\ref{tab:ctrl}. The experimental analysis included identification of the pingers' pulses, estimation of bearing angles and slant delay over time, and application of the clustering solution for two sources.
	\begin{figure*}
		\centering
		\begin{subfigure}[b]{0.4\textwidth}
			\centering
			\includegraphics[width=\textwidth]{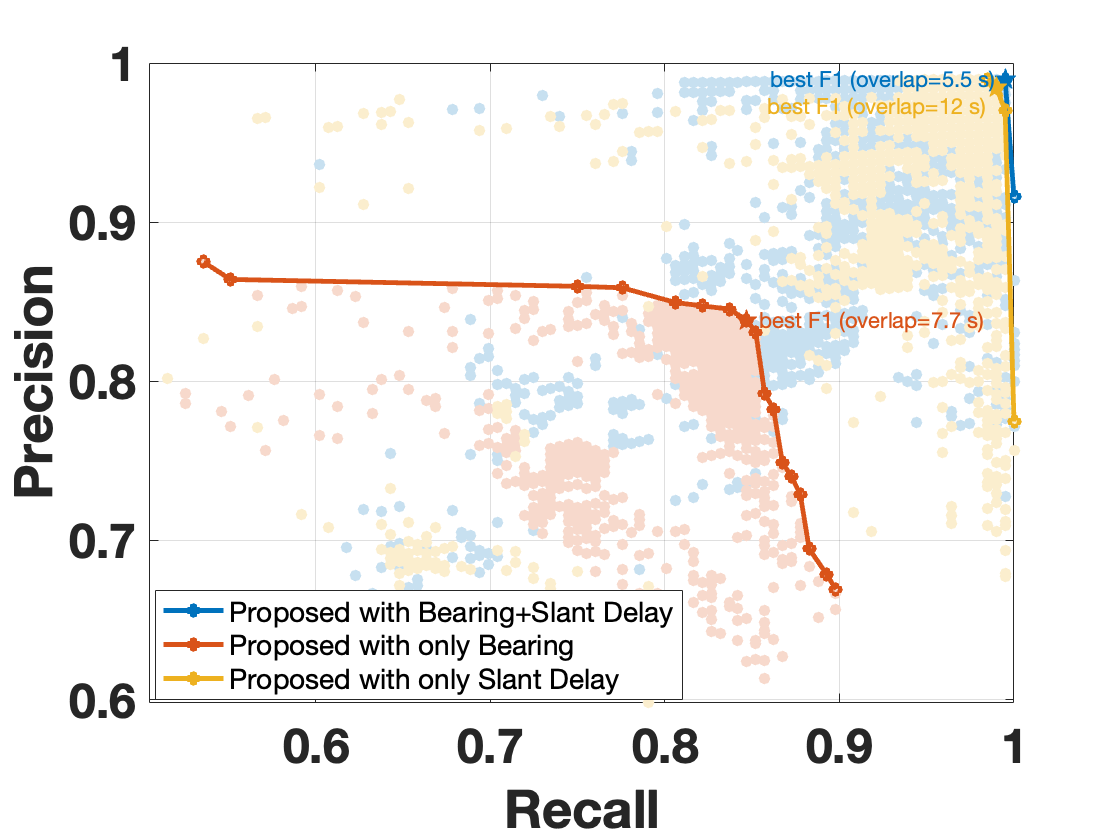}
			\caption{ }
			\label{fig:control_PRcurves}
		\end{subfigure}
		\begin{subfigure}[b]{0.4\textwidth}
			\centering
			\includegraphics[width=\textwidth]{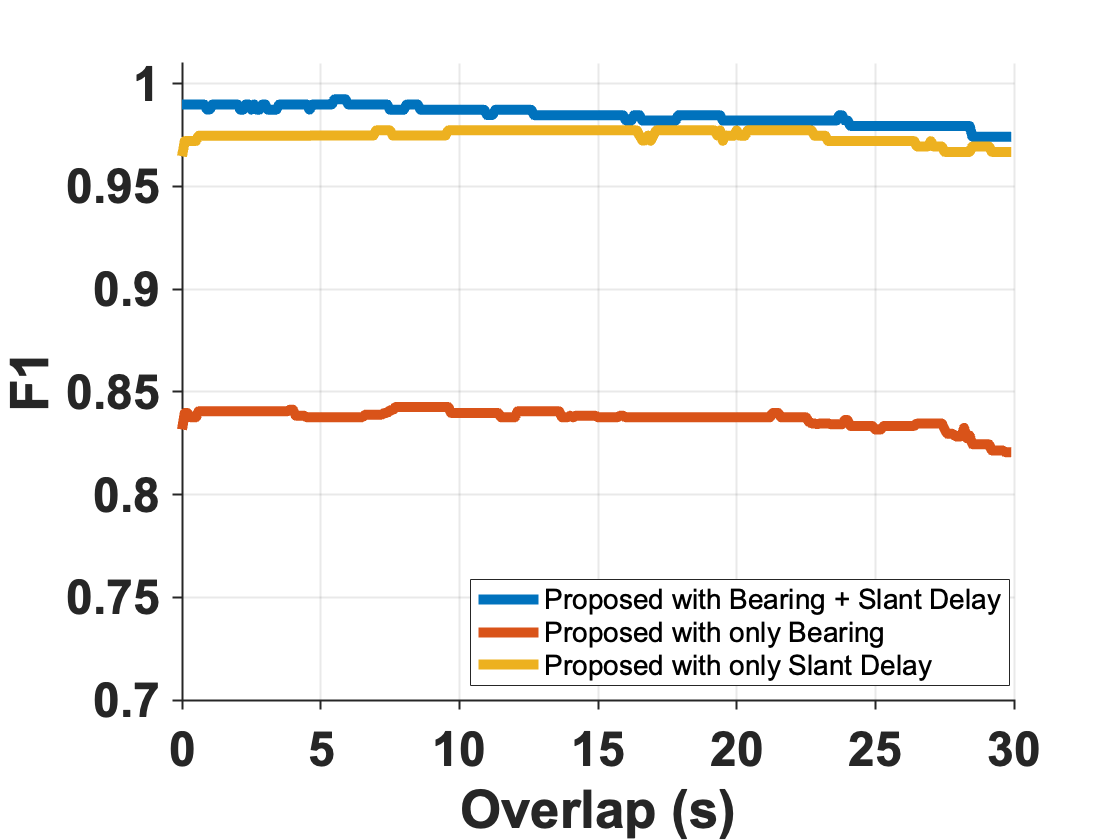}
			\caption{ }
			\label{fig:control_F1_by_overlap_curve}
		\end{subfigure}
		\caption{Sensitivity of the controlled-experiment separation to the overlap parameter. Panel (a) shows the precision–recall operating points obtained by sweeping the overlap. Panel (b) shows F1 as a function of the overlap.}
		\label{fig:control_PRcurves_overlap_curve}
	\end{figure*}
	
	\subsubsection{Experiment Results}
	Performance for various choices of the overlap parameters is shown in Fig.~\ref{fig:control_PRcurves_overlap_curve}. Fig.~\ref{fig:control_PRcurves} shows the precision—recall operating points obtained by sweeping the overlap parameter. The points for the proposed method with joint features and only slant delay variants cluster in the upper-right corner, achieving high recall and high precision simultaneously. The proposed method with only the bearing variant has lower overall precision and a clear precision drop in the high recall region, indicating that bearing alone is sensitive in this context due to the proximity of the two pingers in azimuth and to local reflections. Fig.~\ref{fig:control_F1_by_overlap_curve} shows the F1 score versus the overlap value. Similar to the simulation results, stability is observed for the choice of our only user-defined parameter. In the following, we use $5.5$~ s as the overlap to process the data.
	
\begin{figure*}
		\centering
		\begin{subfigure}[b]{0.32\textwidth}
			\centering
			\includegraphics[width=\textwidth]{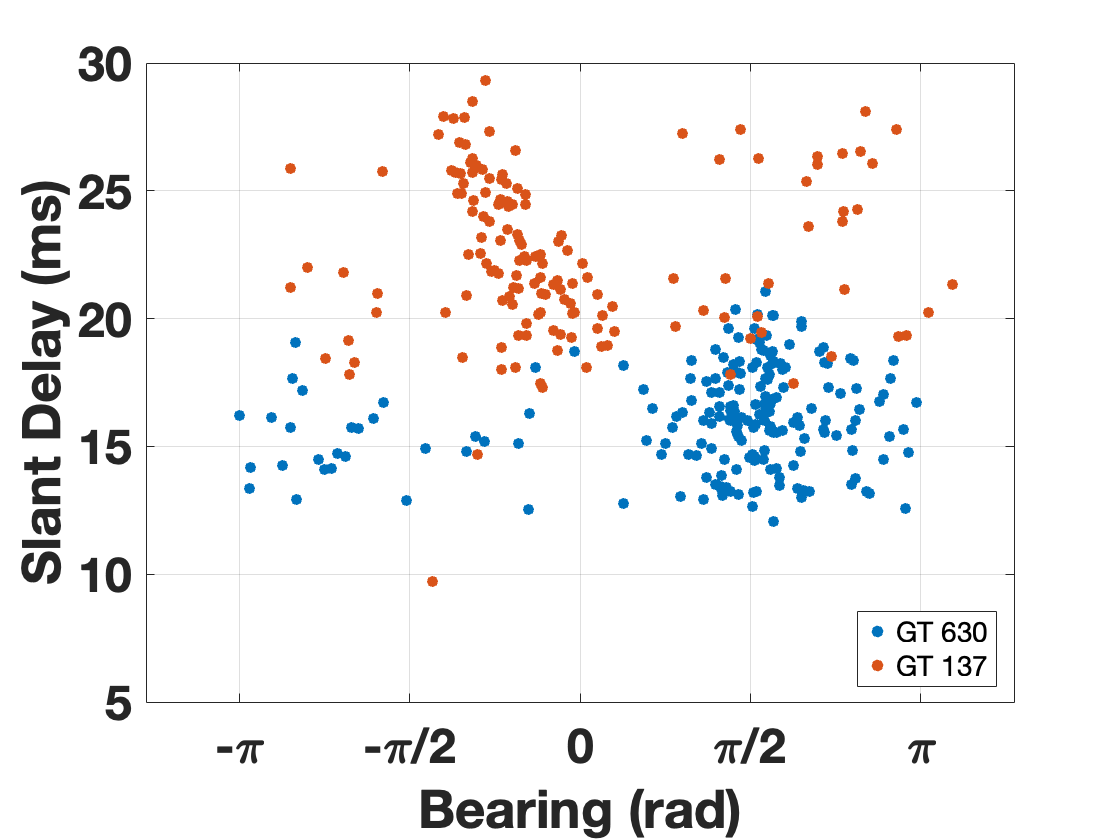}
			\caption{ }
			\label{fig:control_GT}
		\end{subfigure}
		\begin{subfigure}[b]{0.32\textwidth}
			\centering
			\includegraphics[width=\textwidth]{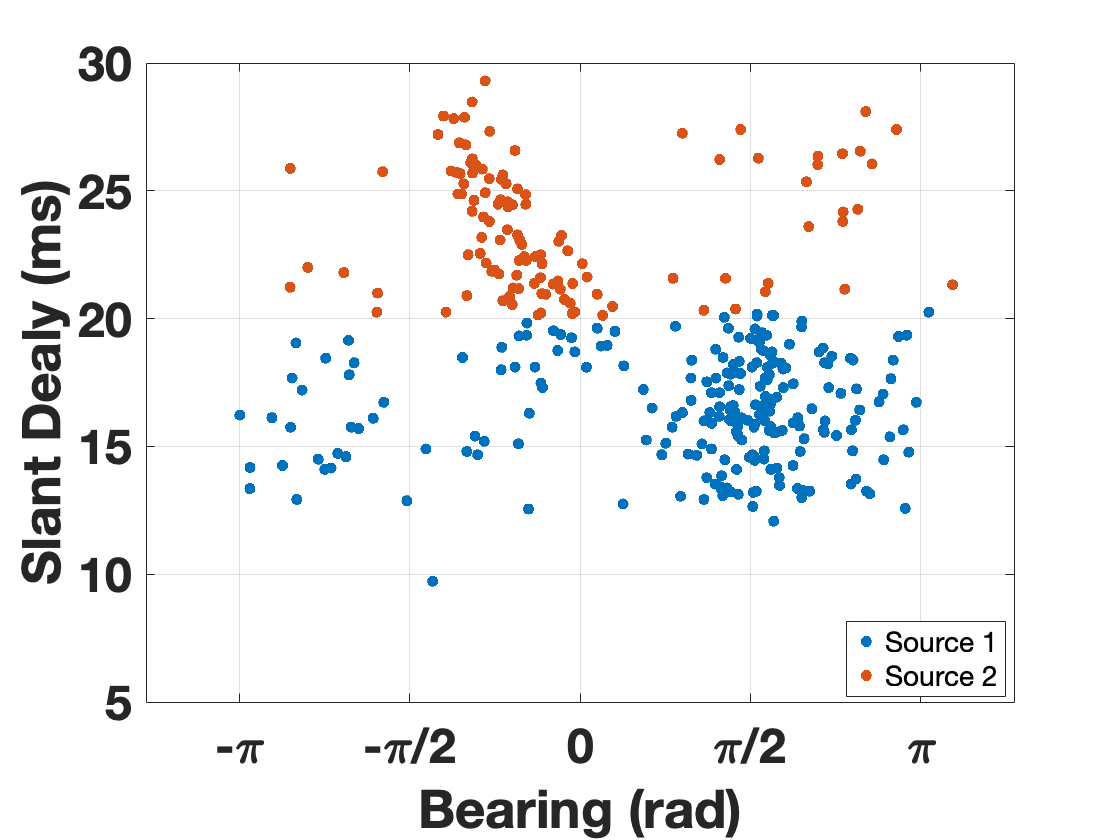}
			\caption{ }
			\label{fig:control_Kmeans_Bearing_DRTD}
		\end{subfigure}
		\begin{subfigure}[b]{0.32\textwidth}
			\centering
			\includegraphics[width=\textwidth]{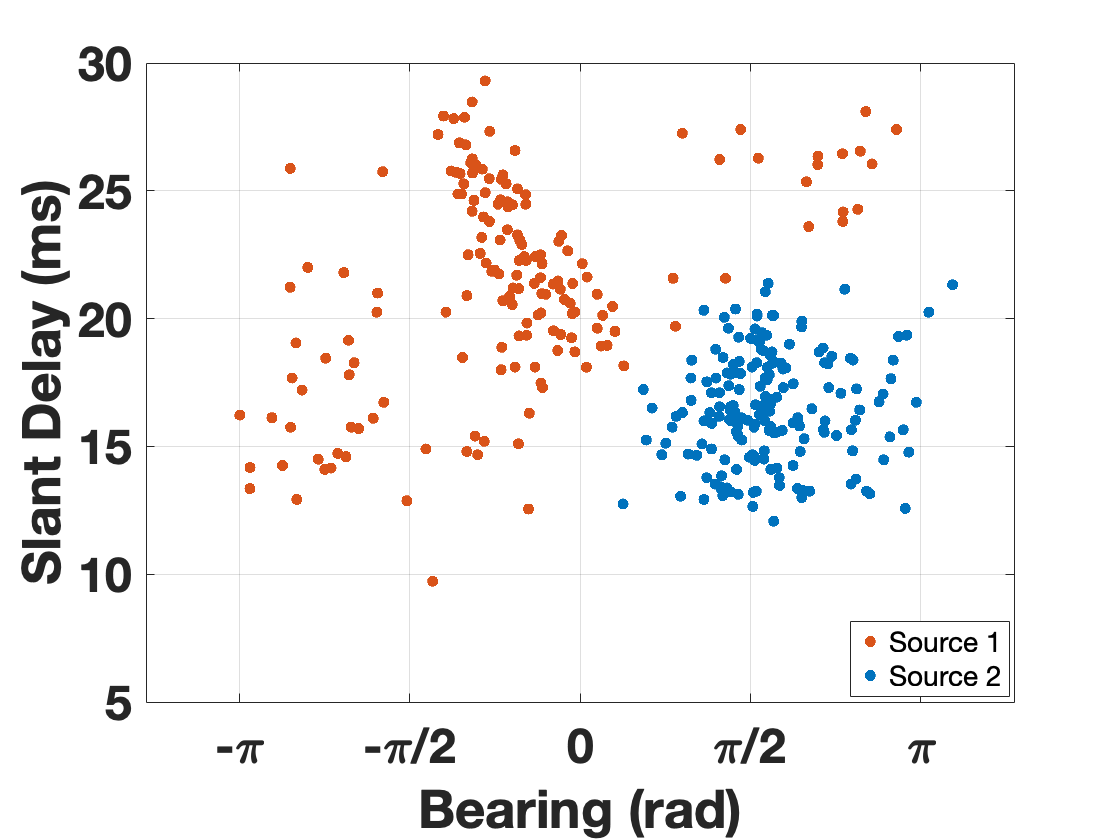}
			\caption{ }
			\label{fig:control_EM_Bearing_DRTD}
		\end{subfigure}
		\begin{subfigure}[b]{0.32\textwidth}
			\centering
			\includegraphics[width=\textwidth]{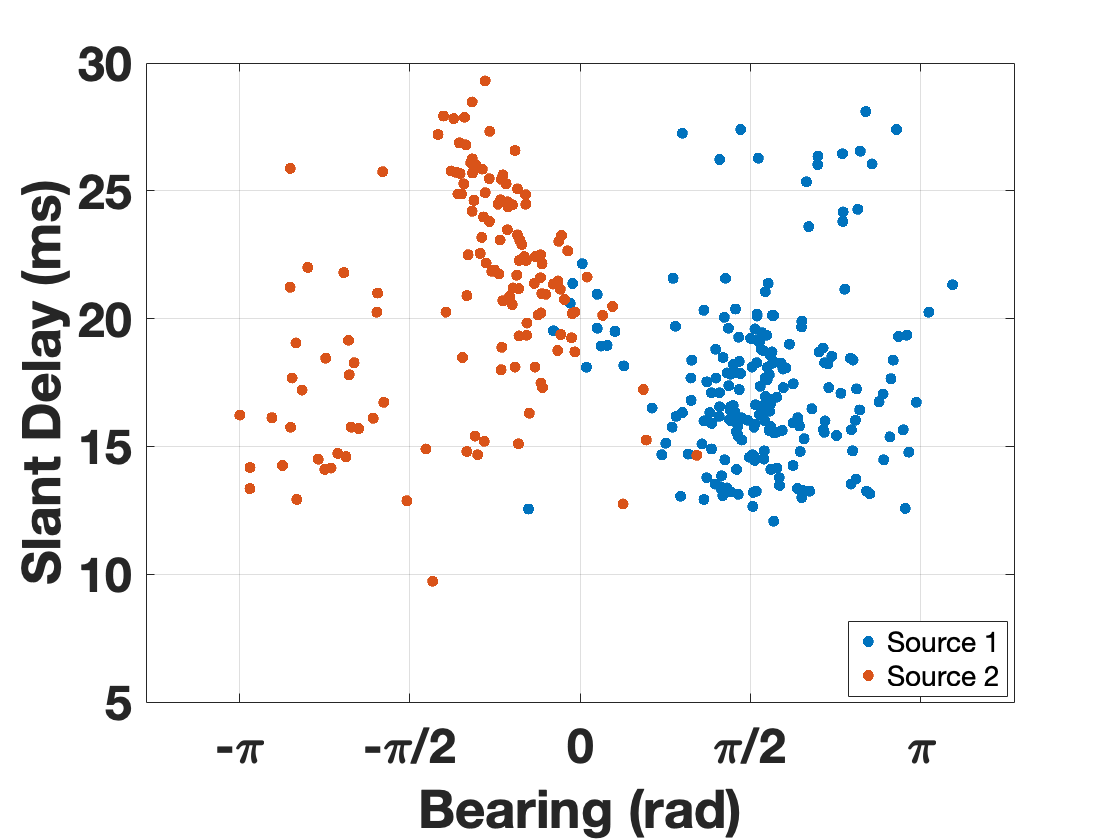}
			\caption{ }
			\label{fig:control_proposed_onlyBearing}
		\end{subfigure}
		\begin{subfigure}[b]{0.32\textwidth}
			\centering
			\includegraphics[width=\textwidth]{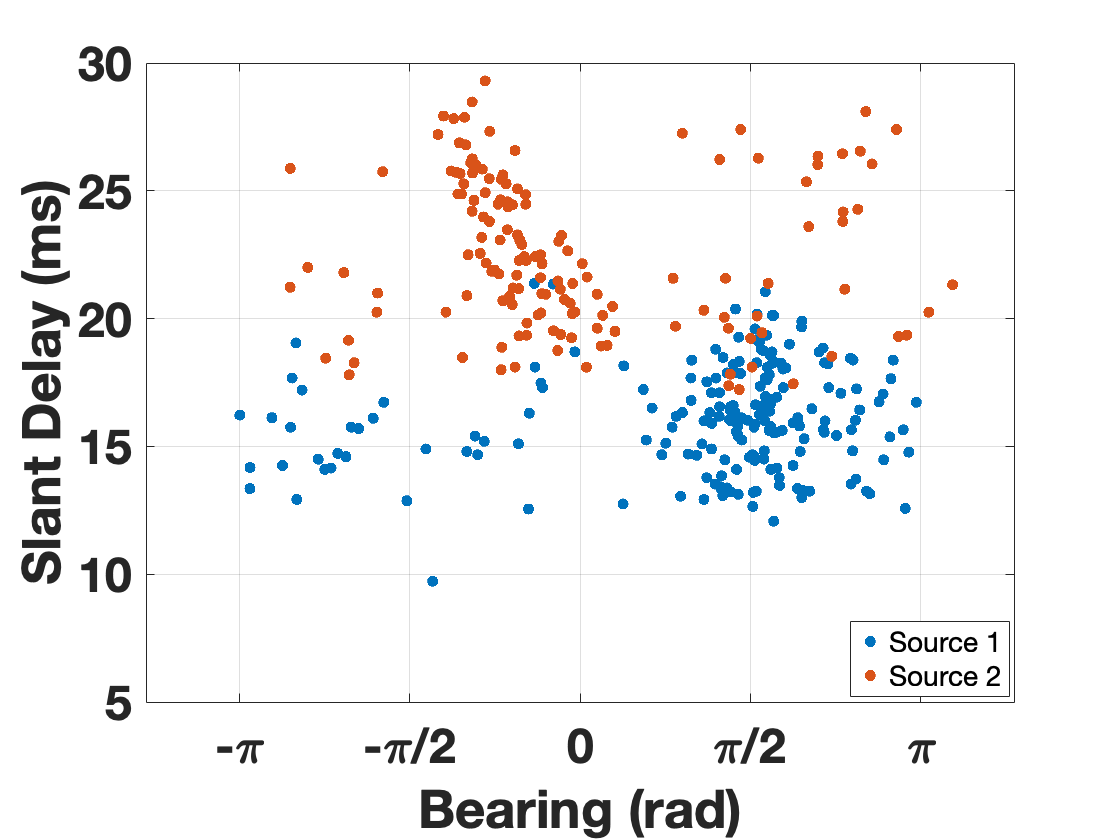}
			\caption{ }
			\label{fig:control_proposed_onlyDRTD}
		\end{subfigure}
		\begin{subfigure}[b]{0.32\textwidth}
			\centering
			\includegraphics[width=\textwidth]{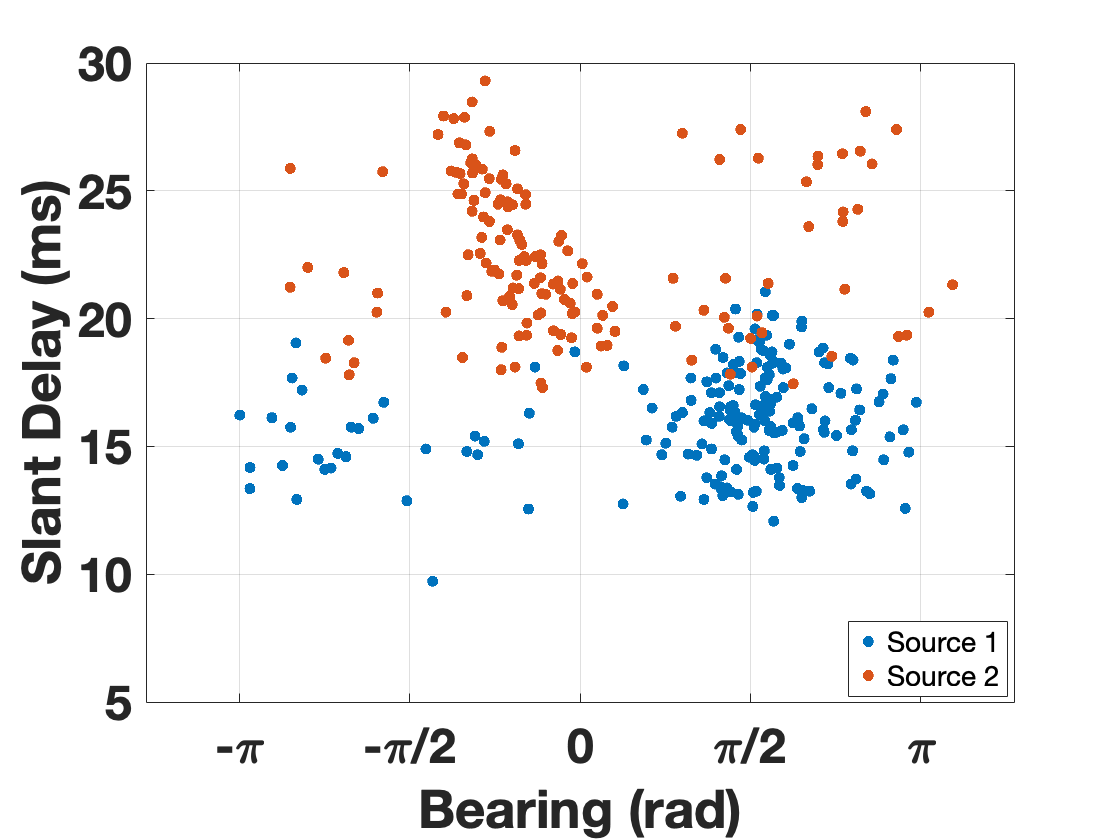}
			\caption{ }
			\label{fig:control_proposed_Bearing_DRTD}
		\end{subfigure}
		\caption{Separation results of the controlled experiment in the feature space at an overlap of $5.5$~s: (a) ground truth, (b) K-means, (c) EM, (d) proposed with bearing only, (e) proposed with slant delay only, and (f) proposed with joint features. The two pingers are globally separable but form banded, curved, and locally crossing distributions rather than two compact clusters, so the static K-means and EM partitions misassign symbols along the boundaries and the curvature. By exploiting temporal continuity across buffers, the proposed variants assign symbols correctly even in the overlap regions, and the joint-feature result in (f) benefits from both cues.}
		\label{fig:control_exp_results}
	\end{figure*}

	Differences between the source assignment solutions of the five considered approaches are shown in Fig.~\ref{fig:control_exp_results} above the feature scatter plot. The ground truth in Fig.~\ref{fig:control_GT} shows two pingers that are globally separable but locally overlapping. The distribution is not composed of two circular clusters but is instead banded, curved, and locally crossing. The K-means scheme in Fig.~\ref{fig:control_Kmeans_Bearing_DRTD} captures the two main centroids and performs reasonably well, but misassigns symbols at the cluster boundaries and along the curvature. The EM in Fig.~\ref{fig:control_EM_Bearing_DRTD} behaves similarly but introduces different errors at the boundaries. This occurs because the true distribution is not strictly elliptical-Gaussian and drifts over time. A common limitation of both EM and K-means is their reliance on the static distribution of the entire record without exploiting temporal continuity. The proposed method with only the bearing variant in Fig.~\ref{fig:control_proposed_onlyBearing} and the proposed method with only the slant delay variant in Fig.~\ref{fig:control_proposed_onlyDRTD} introduce temporal consistency and allow correct assignment even in overlap regions. Results of the joint approach in Fig.~\ref{fig:control_proposed_Bearing_DRTD} appear to benefit from the features of both channels.
	\begin{figure}
		\centering
		\includegraphics[width=0.4\columnwidth]{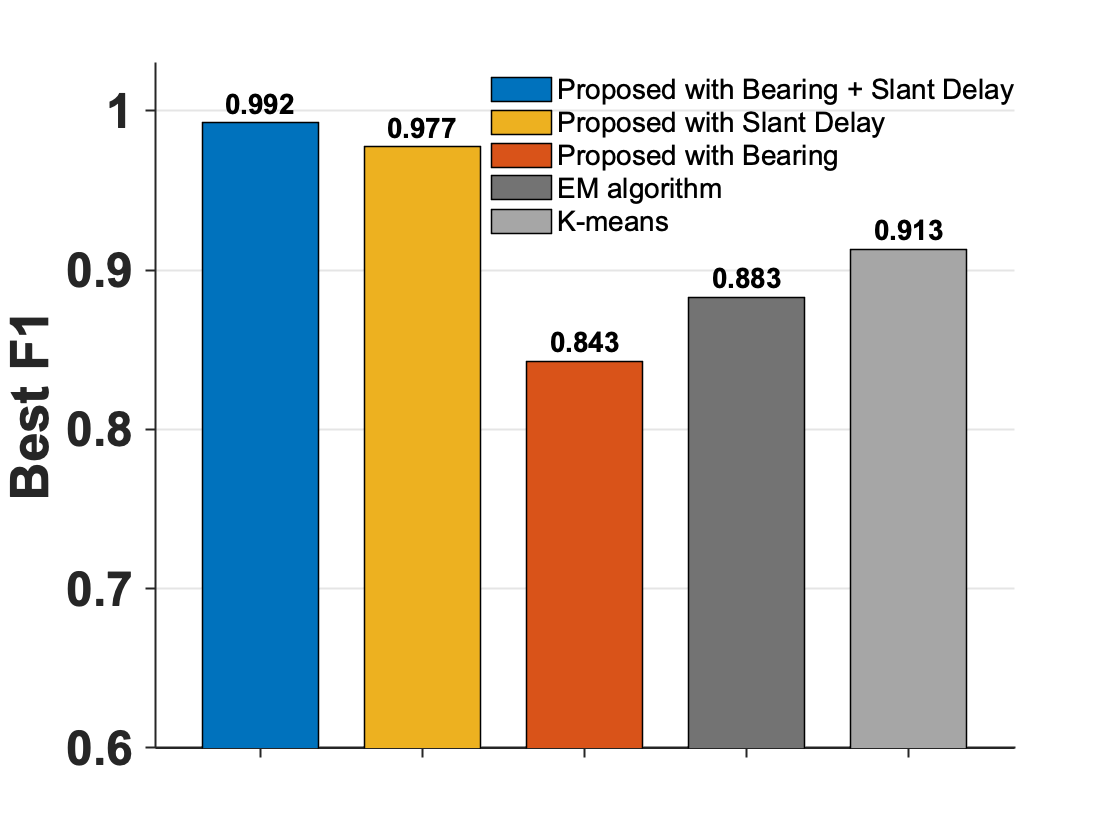}
		\caption{Macro-F1 comparison of the five methods in controlled experiment.}
		\label{fig:control_bestF1_5methods}
	\end{figure}
	
	Ranking by the F1 score is shown in Fig.~\ref{fig:control_bestF1_5methods}. The performance of K-means exceeds that of EM because the two pingers form clearly separated centroids in the two-dimensional feature space. However, K-means still trails the slant delay and joint variants, confirming that a global centroid partition cannot fully address the curvature, local overlap, and temporal drift of the real distribution. In this case, the slant delay is the more discriminative cue, while the vector hydrophone bearing has a residual calibration error. This is mainly due to deploying the vector sensor from a boat that moved rapidly with the waves. Combining the two features still yields the best result, which is precisely why the joint feature is needed. In summary, we observe that the results from the controlled experiment matched those of the numerical simulations, despite the difference in the importance of the bearing and slant delay features.
	
	\subsection{Real Dolphin Whistles Example}
	\begin{figure*}
		\centering
		\begin{subfigure}[b]{0.41\textwidth}
			\centering
			\includegraphics[width=\textwidth]{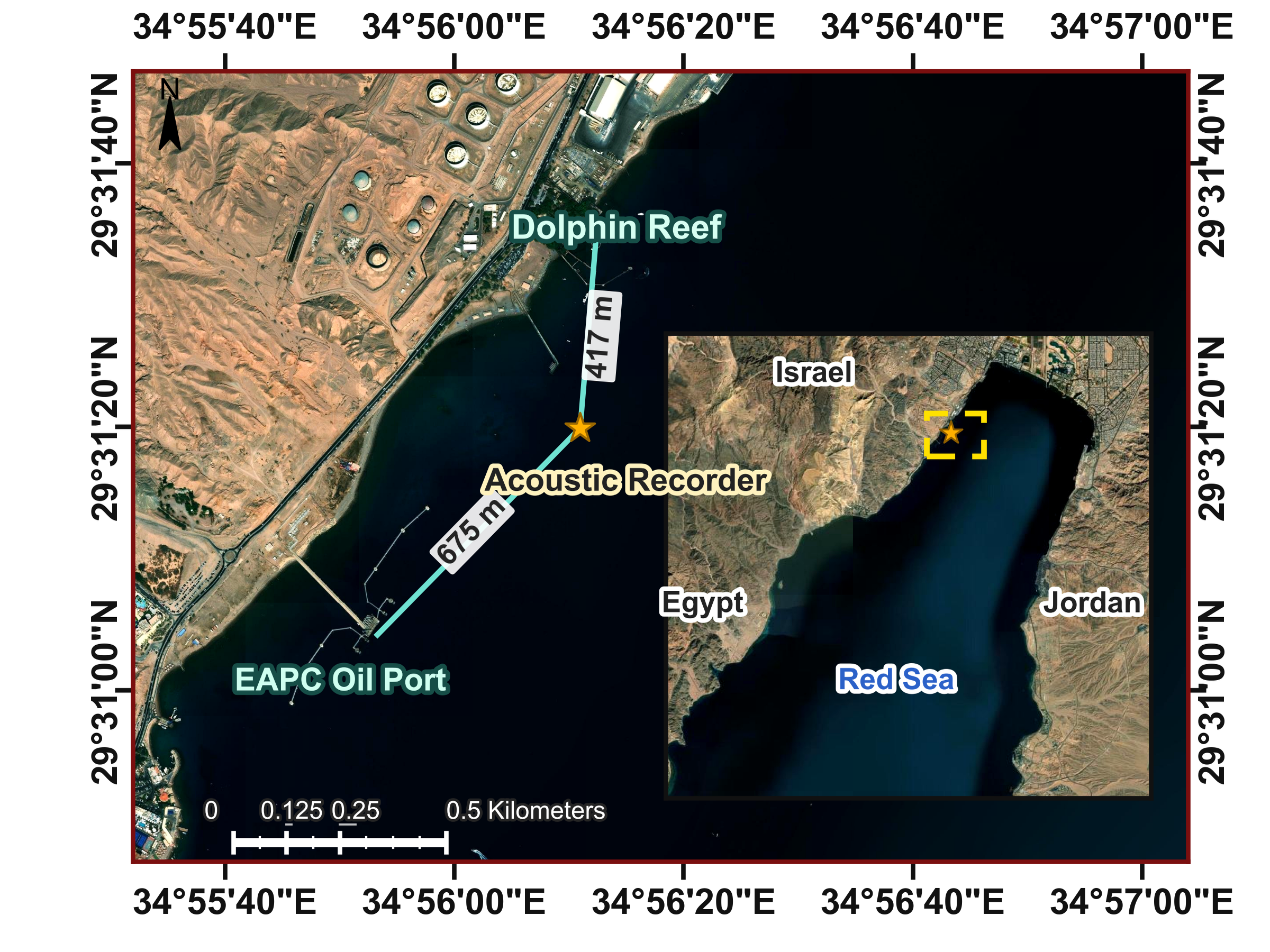}
			\caption{}
			\label{fig:exp_station_map_recreated}
		\end{subfigure}
		\begin{subfigure}[b]{0.45\textwidth}
			\centering
			\includegraphics[width=\textwidth]{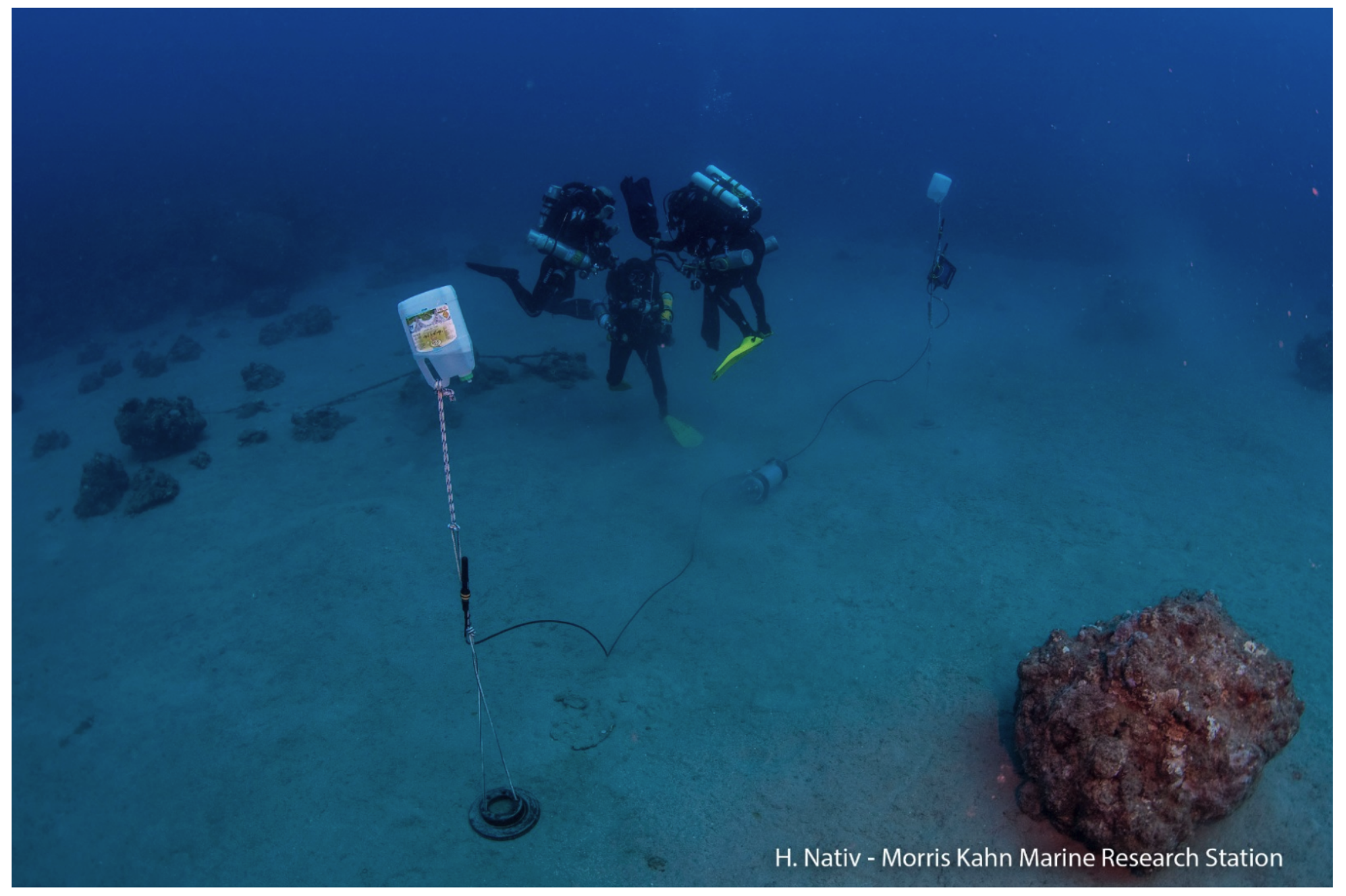}
			\caption{}
			\label{fig:exp_hydrophone_setup_underwater}
		\end{subfigure}
		\caption{Real dolphin whistles data deployment at the Eilat Dolphin Reef, Red Sea. (left) Map of deployment site, (right) underwater two hydrophone deployment setup, which the picture taken from \cite{diamant2024observational}.}
		\label{fig:exp_setup}
	\end{figure*}
	\begin{figure*}
		\centering
		\begin{subfigure}[b]{0.32\textwidth}
			\centering
			\includegraphics[width=\textwidth]{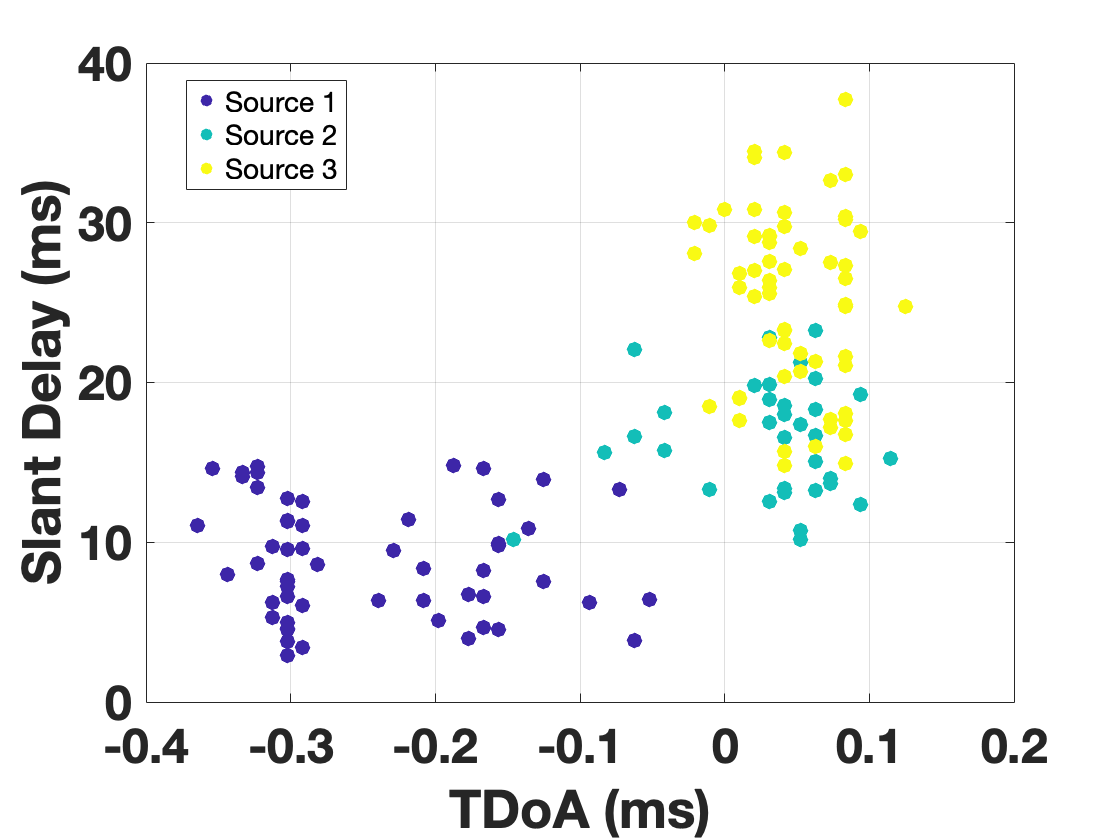}
			\caption{ }
			\label{fig:exp_proposed_TDoA_DRTD}
		\end{subfigure}
		\begin{subfigure}[b]{0.32\textwidth}
			\centering
			\includegraphics[width=\textwidth]{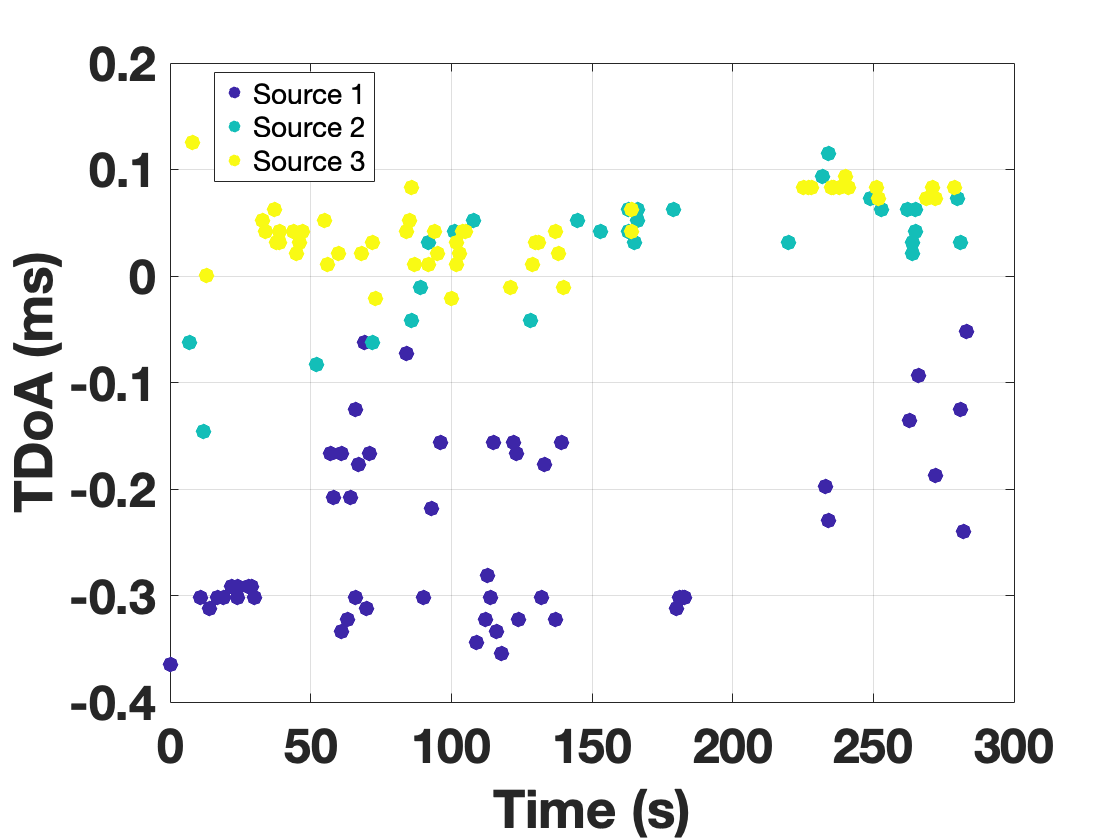}
			\caption{ }
			\label{fig:exp_proposed_time_TDoA}
		\end{subfigure}
		\begin{subfigure}[b]{0.32\textwidth}
			\centering
			\includegraphics[width=\textwidth]{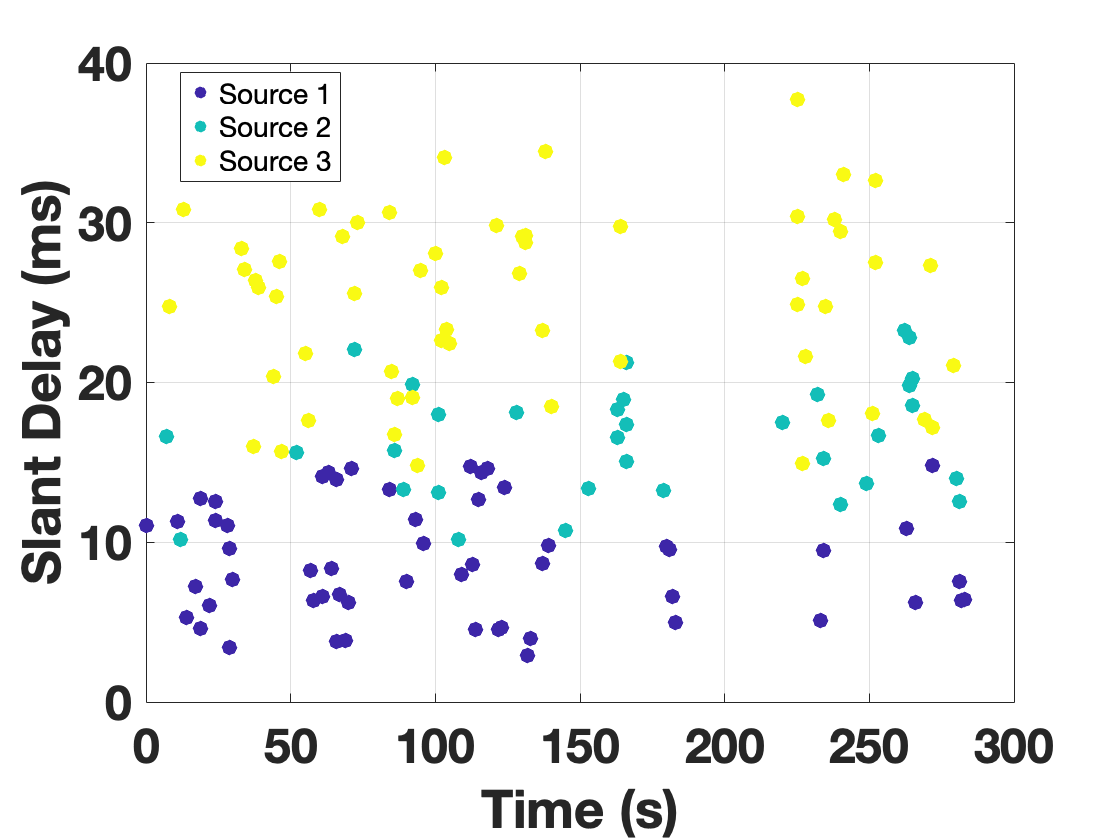}
			\caption{ }
			\label{fig:exp_proposed_time_DRTD}
		\end{subfigure}
		\caption{Real dolphin whistles example separation results in feature space of TDoA and Slant Delay, TDoA in terms of time, Slant Delay in terms of time.
        (a) scatter plots in the joint feature space of TDoA and Slant Delay; (b) TDoA as a function of time; and (c) Slant Delay as a function of time. Both TDoA and Slant Delay are expressed in terms of time.}
		\label{fig:exp_scatter_results}
	\end{figure*}
	The final part of our performance evaluation presents a demonstration of our clustering solution using real dolphin whistles. Our receiver was deployed in Eilat, in the Red Sea, at a depth of 50 m next to an active Dolphin Reef, where three dolphins are free to enter and leave. A full description of this testbed is provided in \cite{diamant2024observational}. In this setup, the receiver consisted of a two-element array with two hydrophones placed 5 m apart and sampled synchronously. Offline detection of the dolphins' whistles followed the procedure described in \cite{korkmaz2023automated}, and estimation of the slant delay feature used the algorithm described in \cite{mishali2024interception}. Without a vector sensor, we used the TDoA feature rather than bearing estimation for these results. Fig.~\ref{fig:exp_setup} shows the deployment site. 
	
	Unlike our previous analysis, since there is no ground truth for which dolphins emit each whistle, we do not quantify the success of the source assignment. A possible way to obtain such ground truth is to deploy tags with acoustic monitoring (e.g., the DTags) on the dolphins and synchronize their emissions with the recorder. However, this requires ethical approval which we could not obtain for wild-living dolphins. However, we report that visual sightings during data collection showed complex dynamics of dolphins, including occlusions. Instead, we present the feature-space structure and temporal consistency to illustrate how the method is applied in a real PAM scenario.
	
	Fig.~\ref{fig:exp_scatter_results} shows the scatter plot of the slant delay and TDoA features along with the clustering solutions for K-means, EM, and the proposed method. In the joint feature space, we observe different assignment solutions. Specifically, with our proposed method, Source~1 is assigned to negative TDoA and low slant delay, Source~2 to near-zero or slightly positive TDoA and intermediate slant delay, and Source~3 to high slant delay. The three groups are not fully separable, with some local overlap at intermediate slant delay levels. While clustering accuracy cannot be evaluated, the fact that the method forms three source clusters in the real data, and that three dolphins were present at the time of recording, indicates the suitability of our approach compared to the baseline.
	
	\section{Discussion}
	\label{sec:discussion}
	Our clustering solution for evaluating the source of dolphin whistles offers a tool for biodiversity analysis and behavioral exploration. The solution enables the assignment of each whistle to a source, allowing us to examine dolphin mobility and social interactions. In particular, changes in bearing/TDoA and slant delay features reveal dynamic changes in dolphin behavior. The number of whistles associated with a single source provides insight into dominance in vocal interactions. Interleaved whistles from different dolphins help us begin to understand the complexity of their interactions. 
    % This approach may offer a new perspective on understanding dolphin community life.
	
	Our joint feature method incorporates channel features rather than dolphins' vocalizations. This reduces reliance on prior information about the dolphins and focuses on changes in their spatial information as the sole basis for source discrimination. Especially when dolphins emit frequent whistles, the proposed method captures temporal continuities in the channel features. Our results show that these temporal relationships are much more important than simply separating the channel feature values. This was demonstrated by the consistent advantage of the proposed approach over common clustering methods that rely on differences in feature distributions. Relying on temporal continuities also makes the proposed method robust across different sea environments (four tested in our analysis) without requiring parameter calibration. The importance of the channel features considered was shown to vary: 
    % bearing errors significantly limit the usefulness of this information, while TDoA information is a more stable discriminating factor. 
    slant delay proved to be an important feature when there are differences in the depths of the source dolphins.
	
	A surprising observation is that, regardless of whether the dolphins move slowly or quickly, a more complex propagation environment benefits our method rather than hinders it. In the shallow San Diego bathymetry with strong relief, the reflection paths create a rich spatial structure in the slant delay, making it highly discriminative. The method then remains nearly flat across overlap, is almost insensitive to the motion pattern, and degrades only slightly with measurement noise. In the deeper and flatter Taiwan bathymetry, all methods perform worse and the curves are steeper. Multipath, usually considered a nuisance in localization, here acts as an environmental fingerprint that the slant delay encodes. This suggests the method is best suited to structured coastal and shelf environments, which matches the environments dolphins usually inhabit.
	
	% A practical benefit of the proposed approach is that the framework requires no signature whistle catalog, no species prior, and no preset number of sources. The method scales nearly linearly with the number of detected whistles, making it well suited to the large, unlabeled datasets typical of long-term PAM. 
 %    % Once whistles are assigned to sources, the temporal evolution of the bearing, TDoA, and slant delay of each source describes how that source moves relative to the receiver. 
 %    In testing with real dolphin recordings, the method formed three source groups, consistent with the three animals known to be present, indicating that the assignments are biologically plausible even without ground truth. We claim that these are relational acoustic quantities rather than calibrated behavioral measures. Mapping them onto ethological constructs such as dominance or individual identity would require independent validation.
	
	We also observe limitations in our approach. First, the method requires a local stationarity assumption. Fast or sharply turning motion accelerates cross-window drift, and when stationarity is violated, the per-buffer channel similarities and their matching degrade. For example, the macro-F1 drops in the FL scenario in the deeper and flatter environment. Second, bias in the measurement noise is not compensated for, unlike the feature distribution reference methods. Third, the proposed method performs symbol separation rather than end-to-end separation, making our scheme more sensitive to detector false alarms and whistle contour errors. 
    % Lastly, our clustering solution is sensitive to false detection of whistles and may classify a wrongly identified noise transient to be another dolphin source. This shortcoming can be mitigated by an outlier detection scheme over the channel feature point cloud at the cost of decreased number of samples. 
    The most significant drawback to address in future work is the reliance on local stationary, which makes our method limited us to small pods (around 10). Replacing the hard per-buffer order selection with an explicit state model would allow drift to be modeled rather than merely tolerated. Additionally, coupling the detector and contour extractor with the clustering stage would enable detection from data and clustering of individual whistle traces into unique whistles. Such extensions would improve the consistency of source assignment, leading to more accurate whistle separation and enhancing the use of passive acoustics for understanding dolphin communication.
	
	\section{Summary and Conclusions}
	\label{sec:conclusions}
	
	In this paper, we present what we believe is the first unsupervised framework for separating dolphin whistles by source without prior knowledge of signature whistles, whistle features, or species. The method clusters the bearing or TDoA and slant delay channel features within short buffers, links the clusters across buffers using Hungarian assignment, and fuses overlapping buffers with reliability weighting. Combining buffer-wise local clustering of bearing or TDoA and slant delay with cross-buffer Hungarian assignment enables near-linear scaling with the number of detected whistles. Results from numerical simulations, controlled experiments, and analysis of real dolphin recordings demonstrate that our method outperforms common blind clustering solutions such as K-means and EM discriminators in terms of sensitivity to measurement noise, robustness to motion patterns, and number of sources. The method uses only a single user-defined parameter, which contributes to its robustness. This robustness was demonstrated in four different marine environments. These results indicate the benefit of combining geometric and temporal discriminators for long-term PAM analysis.

%% Author-identifying declarations and acknowledgments are provided in the
%% separate ESWA title page file, as required for double-anonymized review.

\balance
%% IEEE numbered bibliography using the original complete database.
\bibliographystyle{IEEEtranDOI}
\bibliography{refs}

\end{document}